\documentclass[]{aa} 
\usepackage[output-decimal-marker={.}, exponent-product=\cdot, per-mode=fraction]{siunitx}
\usepackage{xcolor}

\newcommand{\abs}[1]{\left| #1 \right|}

\begin{document}

\title{Photospheric magnetic structure of coronal holes}

\author{Stefan J. Hofmeister\inst{\ref{inst1}} \and Dominik Utz\inst{\ref{inst1}} \and Stephan G. Heinemann\inst{\ref{inst1}} \and Astrid Veronig\inst{\ref{inst1}, \ref{inst2}} \and Manuela Temmer\inst{\ref{inst1}}}

\institute{Institute of Physics, University of Graz, Austria\label{inst1} \and 
Kanzelh\"ohe Observatory for Solar and Environmental Research, University of Graz, Austria\label{inst2}}

\abstract{
In this study, we investigate in detail the photospheric magnetic structure of 98 coronal holes using line-of-sight magnetograms of SDO/HMI, and for a subset of 42 coronal holes using HINODE/SOT G-band filtergrams. We divided the magnetic field maps into magnetic elements and quiet coronal hole regions by applying a threshold at \SI{\pm 25}{G}. We find that the number of magnetic bright points in magnetic elements is well correlated with the area of the magnetic elements (cc=$0.83 \pm 0.01$). Further, the magnetic flux of the individual magnetic elements inside coronal holes is related to their area by a power law with an exponent of $1.261 \pm 0.004$ (cc=$0.984 \pm 0.001$). Relating the magnetic elements to the overall structure of coronal holes, we find that on average \SI{69 \pm 8}{\percent} of the overall unbalanced magnetic flux of the coronal holes arises from long-lived magnetic elements with lifetimes \SI{>40}{hours}. About \SI{22 \pm 4}{\percent} of the unbalanced magnetic flux arises from a very weak background magnetic field in the quiet coronal hole regions with a mean magnetic field density of about \SIrange{0.2}{1.2}{G}. This background magnetic field is correlated to the flux of the magnetic elements with lifetimes of \SI{>40}{h} (cc=$0.88 \pm 0.02$). The remaining flux arises from magnetic elements with lifetimes \SI{<40}{hours}. 
By relating the properties of the magnetic elements to the overall properties of the coronal holes, we find that the unbalanced magnetic flux of the coronal holes is completely determined by the total area that the long-lived magnetic elements cover (cc=$0.994 \pm 0.001$). 
}

\maketitle

\section{Introduction}

Coronal holes are known as large-scale structures in the solar corona characterized by a reduced temperature, density, and an "open" to interplanetary space magnetic field topology. Along the open magnetic field lines rooted in the coronal holes, plasma is accelerated to supersonic speeds forming interplanetary high-speed solar wind streams transcending our solar system. These plasma streams shape the interplanetary space and cause geomagnetic storms whenever they sweep over the Earth.

Although coronal holes were already discovered in the 1973-1974 Skylab epoch \citep{tousey1973, krieger1973, glencross1974},  the source of their open magnetic flux still remains unclear. It is still under debate if coronal holes are magnetically rooted deep in the convection zone as suggested by their observed differential rotation rates, if their open magnetic flux is induced by the outflowing solar wind plasma, or if their open magnetic flux simply consists of ordinary closed quiet-Sun loops that are opened up by some mechanisms. But, before we can decide what the source of the open magnetic flux of coronal holes is, we have to determine which of the many kinds of photospheric magnetic substructures actually contribute to the open magnetic flux.

From quiet Sun high-resolution observations, we know that the vertical photospheric magnetic flux is distributed in a salt-and-pepper manner. Surface flows from convective motions transport magnetic flux from the inner toward the outer regions of supergranular cells, resulting in localized magnetic field concentrations at their edges; these concentrations are known as magnetic elements. These are randomly distributed  in the supergranular lanes over the solar surface with positive and negative polarities resulting in the salt-and-pepper picture \citep{lites2008}. The distribution of lifetimes of magnetic elements splits them into three categories, whereby their half-lives match the timescales of convective motions, i.e., granulation, meso-granulation, and supergranulation. 
The magnetic elements themselves are built up of multiple small-scale magnetic fibers, which are small-scale, kilo-Gauss strong solar magnetic field concentrations observed in the photosphere and chromosphere \citep{berger2004,liu2018}. Since their field strength exceeds the equipartition field strength in the photosphere of \SI{\approx 500}{G}, they are thought to be created by convective collapse in the intergranular lanes \citep{spruit1979, nagata2008, utz2014}. When the small-scale strong magnetic fibers are aligned to the line of sight (LoS), their cross section can be well observed as bright points in photospheric G-band filtergrams. 
These magnetic bright points (MBPs) are of scientific interest owing to many processes in which they are involved, such as magnetohydrodynamics wave creation and propagation \citep[e.g.,][]{kato2011,mathioudakis2013}.

In contrast, the inner of supergranular cells show much less and weaker vertical magnetic field regions than their edges; they are more covered by horizontal magnetic fields on mesoscales belonging to closed loops. The vertical magnetic field concentrations present only have mean magnetic flux densities of a few \SI{100}{G}, i.e., below the equipartition field strength, and thus represent a distinctly different phenomena than their counterparts in the supergranular lanes \citep{lites2008}.

When we compare the photospheric magnetic field of coronal holes and quiet Sun regions at the \SI{1.1}{arcsec} resolution of the Helioseismic and Magnetic Imager (HMI) on board the Solar Dynamics Observatory (SDO), they look very similar. Most of the inner parts of supergranular cells seem to be almost flux-free of vertical magnetic fields, whereas the magnetic flux is concentrated in magnetic elements in the supergranular lanes.
However, differences were found in the three-dimensional magnetic field topology. In coronal, holes, the magnetic elements are the footpoints of open-to-interplanetary-space magnetic funnels, i.e., special clusters of vertical magnetic fibers, which are thought to be the small-scale source regions of high-speed solar wind streams \citep{hassler1999, tu2005} and contain \SI{80}{\percent} of the unbalanced magnetic flux of the coronal holes \citep{hofmeister2017, heinemann2018}. Further, the closed loops within coronal holes are almost all low-lying not even reaching the corona, whereas in quiet Sun regions these closed loops entirely cover the corona \citep{wiegelmann2004}. 

Therefore, the differences between coronal holes and quiet Sun regions are the open magnetic funnels that shape the three-dimensional magnetic topology. However, since both open magnetic funnels and the closed loops are rooted in photospheric magnetic elements which cannot be easily distinguished, their difference has never been investigated in detail. Consequently, which magnetic elements in fact contribute to the unbalanced magnetic flux of coronal holes and how they set the magnetic properties of the overall coronal hole has never been studied. This is the purpose of this study.

This comprehensive study is structured as follows: After describing the datasets and methods (Sect. \ref{secdata}), we start with a qualitative investigation of the photospheric magnetic field within coronal holes (Sect. \ref{sect_visual}). Then, we analyze the individual properties of the magnetic elements and study their distribution in coronal holes dependent on their life times (Sect. \ref{sec_structure}). Thereby, we are able to determine which kinds of the magnetic elements are the source of most of the unbalanced magnetic flux of coronal holes, i.e., which kinds are responsible for the open magnetic funnels, and which belong to the closed loops. Then, we analyze the distribution of the magnetic field in the quiet coronal hole regions, i.e., the coronal hole region outside of the magnetic elements, which are still responsible for \SI{\approx 20}{\percent} of their unbalanced magnetic flux (Sect. \ref{sec_quietch}). Following, we demonstrate how the overall magnetic properties of coronal holes, i.e, their mean signed and unsigned magnetic field strength and its ratio, i.e., the percentaged unbalanced magnetic flux, are defined by the distribution of the magnetic elements and the magnetic field in quiet coronal hole regions (Sect \ref{sec_overall}). Finally, we discuss and review our results (Sect. \ref{sec_discuss} and \ref{sec_conclude}).

\section{Datasets and methods} \label{secdata}
We identify the coronal holes in the \SI{193}{\angstrom} filtergrams taken by the Athmospheric Imaging Assembly (AIA) onboard SDO. These filtergrams monitor the emission of Fe XII ions in the solar corona at a temperature of $\approx \SI{1.6e6}{K}$ (peak response). They are especially well suited to extract coronal holes because of their high contrast between coronal hole and quiet Sun regions. The spatial resolution of these filtergrams is \SI{1.5}{arcsec} at a plate scale of \SI{0.6}{arcsec/pixel} and their temporal resolution is \SI{12}{seconds} \citep{lemen2012}. From August 2010 to January 2017, we extracted the coronal holes from these filtergrams using a thresholding technique based on \citet{rotter2012} at a cadence of one image per day. Hereby, whenever it is applicable, we extract the coronal holes at the exact time when the center of the Sun was also observed by the Solar Optical Telescope (SOT) on board Hinode. From this dataset, we select those coronal holes that have their center of mass located near the central meridian and whose complete area is within the distance of \SI{35}{\degree} around the solar disk center to minimize projection effects. These criteria result in a final dataset of 98 coronal holes. We calculate for each coronal hole its projection-corrected area $A_\text{CH}$, and latitude $\varphi_\text{CH}$ and longitude $\lambda_\text{CH}$ of its center of masses. Details on the computations are described in \citet{hofmeister2017}.

The photospheric magnetic field below coronal holes is analyzed using the SDO/HMI  LoS magnetograms as derived by the HMI Doppler camera, which observes the Zeeman splitting of photospheric Fe I ions at \SI{6173.3}{\angstrom}. This data product contains high-cadence, low-noise magnetograms at a spatial resolution of \SI{0.91}{arcsec} at a plate scale of \SI{0.505}{arcsec/pixel} with a cadence of \SI{45}{seconds} and a photon noise level of \SI{\approx 7}{G} near the disk center \citep{scherrer2012, schou2012, couvidat2016}. Since we aim to analyze small-scale magnetic features within coronal holes, we use the magnetograms at full resolution, meaning that we do not rescale the magnetograms to the plate scale of the AIA \SI{193}{\angstrom} images in order to avoid interpolation effects. The mean magnetic field density of each pixel of the magnetograms is further corrected for the assumption of a radial mean magnetic field density following \citet{hofmeister2017}. We project the coronal holes as extracted from the AIA \SI{193}{\angstrom} images to the magnetograms, and derive their photospheric mean magnetic field strength $B_\text{CH}$, unbalanced, i.e., presumably open, magnetic flux $\Phi_\text{CH}$, total unsigned magnetic flux $\Phi_\text{CH,us}$, and percentaged unbalanced magnetic flux $\Phi_\text{CH,pu} = \Phi_\text{CH} / \Phi_\text{CH,us}$. For simplicity, throughout the article, we only give the absolute values of the magnetic properties of the coronal holes. Again, for more details on the computations, we refer to \citet{hofmeister2017}.

We then apply a thresholding technique at \SI{\pm 25}{G} to the magnetograms. We refer to all magnetogram pixels below this threshold as quiet coronal hole regions. The distribution of the magnetogram pixels in the quiet coronal hole regions gives us an estimate on the average noise level and average background magnetic field in the magnetogram within the coronal hole; the corresponding procedures are described in Section \ref{sec_quietch}. For all coronal holes, the noise level is in the range \SIrange{8}{12}{G}  depending on the position of the coronal hole on the solar disk, with a mean noise level of \SI{9.2}{G}. Further, we calculate the unbalanced magnetic flux arising from the quiet coronal hole regions, $\Phi_\text{qu}$. 

All magnetogram pixels above the threshold cluster to small, unipolar regions. Whenever such a cluster exceeds a size of \SI{2x2}{pixels}, we refer to it as a magnetic element. The core size of 2x2 pixels is required to reject spurious false-positive magnetic elements from our dataset that arise from magnetogram noise. This extraction technique was tested for its sensitivity on noise on artificial magnetograms consisting solely of noise, and works stable up to a local noise level of \SI{14}{G}.
Further, the results were visually checked on the real magnetograms. 
For each magnetic element, we derived the position of the magnetic element $\varphi_{me}$ and $\lambda_{me}$, its area $A_{me}$, its mean magnetic field density $B_{me}$, and its magnetic flux $\Phi_{me}$. Further, we counted the number of magnetic elements $N_\text{me}$ within each coronal hole. The magnetic properties of the magnetic elements were then transformed such that positive values represent the dominant coronal hole polarity. We calculated the lifetime $\tau_{me}$  of the magnetic elements by tracking the magnetic elements for \SI{\pm 2}{days}, at a cadence of \SI{45}{seconds} for the first \SI{\pm 3}{hours}, at a cadence of \SI{90}{seconds} for the next \SI{3}{hours}, \SI{180}{seconds} for the following \SI{6}{hours}, and \SI{360}{seconds} for the remaining \SI{1.5}{days}. The tracking was performed by cross-correlating the two consecutive HMI-LoS magnetograms to find the relative shift, applying the shift, and assigning the magnetic elements by checking which magnetic elements in the subsequent magnetogram overlap with the magnetic elements in the preceding magnetogram. In case of "splitting", i.e., that multiple magnetic elements in the subsequent magnetogram overlap with one magnetic element in the preceding magnetogram, only the largest magnetic element in the subsequent magnetogram was tracked.

The fine structures of a subset of the magnetic elements are analyzed using the Hinode/SOT G-band filtergrams. Hinode is a Japanese, US, and European satellite mission \citep{kosugi2007} dedicated to the understanding of solar magnetic fields, especially in regards to small-scale solar magnetic fields. Among the suite of instruments is the SOT \citep[][]{tsuneta2008}, which itself contains three sub-instruments. One of these is the Broad-band Filter Imager (BFI) which takes part in a synoptic program. Within this observation program, Hinode has taken about one image per day for all broadband filters (red, green, and blue continuum as well as G band, Cn band, and Ca II H) near the disk center from the mission start at the end of October 2006 until spring 2016 when a malfunctioning CCD ended the operational life of the BFI and the Narrowband Filter Imager (NFI) sub-instruments of SOT. The field of view (FOV) is about \SI{200x100}{arcsec}, the spatial resolution \SI{0.2}{arcsec} and the plate scale \SI{0.109}{arcsec/pixel}. 

Within the Hinode FOV, we first extracted the photospheric  MBPs by applying an automated image segmentation and identification alogrithm to the G-band filtergrams as described in \citet{utz2009,utz2010}. Then, we co-aligned the SOT filtergrams to the HMI magnetograms by cross-correlating the binary maps of MBPs and magnetic elements and thereby adapting the $x$ and $y$ offsets, the $x$ and $y$ scaling-factors, and the roll-angle of the images. Finally, we assigned the MBPs to the magnetic elements by checking the pixel overlap, and derived the total number of MBPs $N_\text{MBP}$ and the number of MBPs within a magnetic elements $N_\text{MBP,me}$. In total, for 42 of the 98 coronal holes high-resolution G-band observations with a limited FOV have been available.

Since this study involves two different spatial scales, i.e., the scales of magnetic elements and coronal holes, we also used two typical scales for their parameters. For the area and magnetic flux of magnetic elements, we used \SI{e18}{Mx} and \si{Mm^2} as a basis scale, which is the magnetic flux and size of a typical small magnetic element in our dataset. For the number density of magnetic elements in coronal holes, we used \SI{e-4}{Mm^2}, which is the inverse  of the typical size of a small coronal hole. The contribution of magnetic elements to the mean magnetic field strength of coronal holes is given in \si{Mx/cm^2}.
All statistical uncertainties in this study stated give the standard deviation of the parameters derived, the $1\sigma$ uncertainty range of the fit parameters, and the $1\sigma$ uncertainty of the  correlation coefficients, respectively. All correlation coefficients give the Pearson's correlation coefficient unless otherwise specified.

\section{Qualitative investigation of the photospheric magnetic field of coronal holes}
\label{sect_visual}

In this section, we analyze the photospheric magnetic field below coronal holes by studying a typical medium-size coronal hole. It is thought to be a descriptive basis for the following sections, and the results are representative for our entire coronal hole dataset. The coronal hole was observed by SDO on July 25, 2013 near the center of the disk, has a positive magnetic field polarity, and a mean magnetic field strength of \SI{5.4}{G}. 

\begin{figure*}[tp]
\centering
\includegraphics[width = \textwidth]{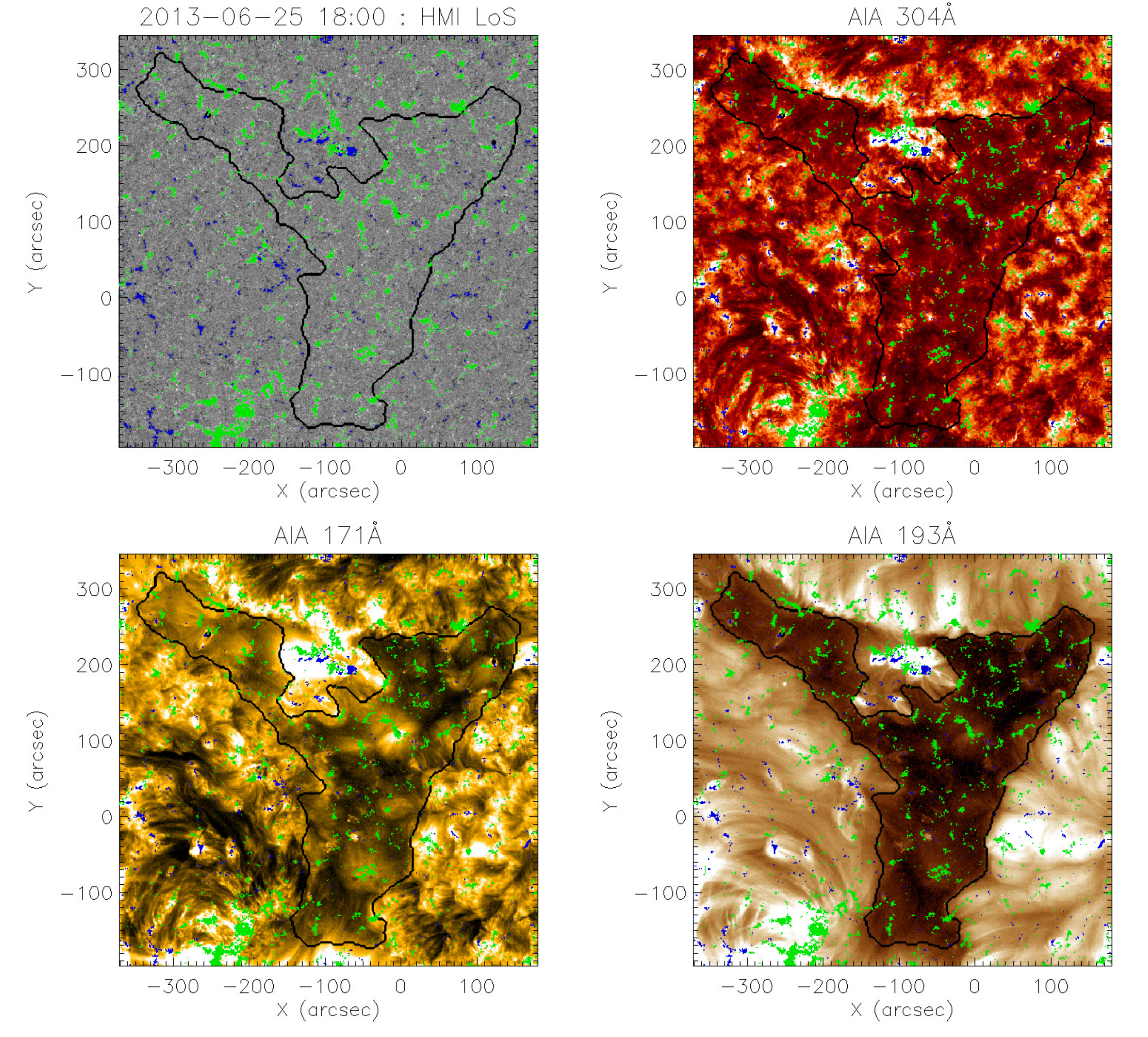}
\caption{From left to right and top to bottom: SDO/HMI LoS magnetogram, AIA \SI{304}{\angstrom}, \SI{171}{\angstrom}, and \SI{193}{\angstrom}  filtergram of the solar chromosphere and corona taken on July 25, 2013. The magnetogram is scaled to \SI{\pm 50}{G}; the AIA channels are sensitive to plasma at a temperature of \SI{50}{kK} (AIA \SI{304}{\angstrom}), \SI{0.63}{MK} (AIA \SI{171}{\angstrom}), and \SI{1.25}{MK}, respectively (AIA \SI{193}{\angstrom}). The green contours outline magnetic elements with dominant magnetic polarity, the blue contours with nondominant polarity.}
\label{corona}
\end{figure*}

In the top left panel of Figure \ref{corona}, we show the photospheric SDO/HMI-LoS magnetogram of the coronal hole scaled to \SI{\pm 50}{G}. The boundary of the coronal hole is overlayed in black, and regions exceeding a magnetic field strength of $>\abs{25}$\,G are outlined in green and blue, respectively. Most of the photospheric area below coronal holes is covered by weak magnetic fields (top left panel, gray). Within these weak magnetic fields, magnetic elements with absolute magnetic field strength \SI{>> 25}{G} are apparent; most of these have a positive magnetic polarity (green) and a few have a negative polarity (blue). It is apparent that these magnetic elements outline the supergranular cells and form the photospheric magnetic network. 
The further three panels of Figure \ref{corona} portray the AIA \SI{304}{\angstrom}, AIA \SI{171}{\angstrom}, and AIA \SI{193}{\angstrom} filtergrams of the coronal hole, showing plasma from chromospheric to coronal temperatures with peak sensitivities at \SI{\approx 50}{kK}, \SI{800}{kK}, and \SI{1.6}{MK}, respectively. We note that the formation of the He II \SI{304}{\angstrom} line contains a contribution from photoionization by coronal back-radiation \citep{zirin1996}; therefore an imprint of the coronal hole is visible in the upper chromospheric He II line. 
Further, we do not expect substantial projection effects from the photosphere to the corona since the position of the coronal hole is near the center of the solar disk. From these images, we find that the darkest coronal regions are in general located above the inner parts of the magnetic network cells. Thus, they are not located above strong magnetic elements with the dominant, positive coronal hole polarity, but are surrounded by these elements. In contrast, the brightest coronal features within the coronal hole are located close to magnetic elements with nondominant, negative coronal hole polarity, relating these features to coronal bright points, small-scale loops, and jets.

\begin{figure*}[tp]

{\includegraphics[width=\textwidth]{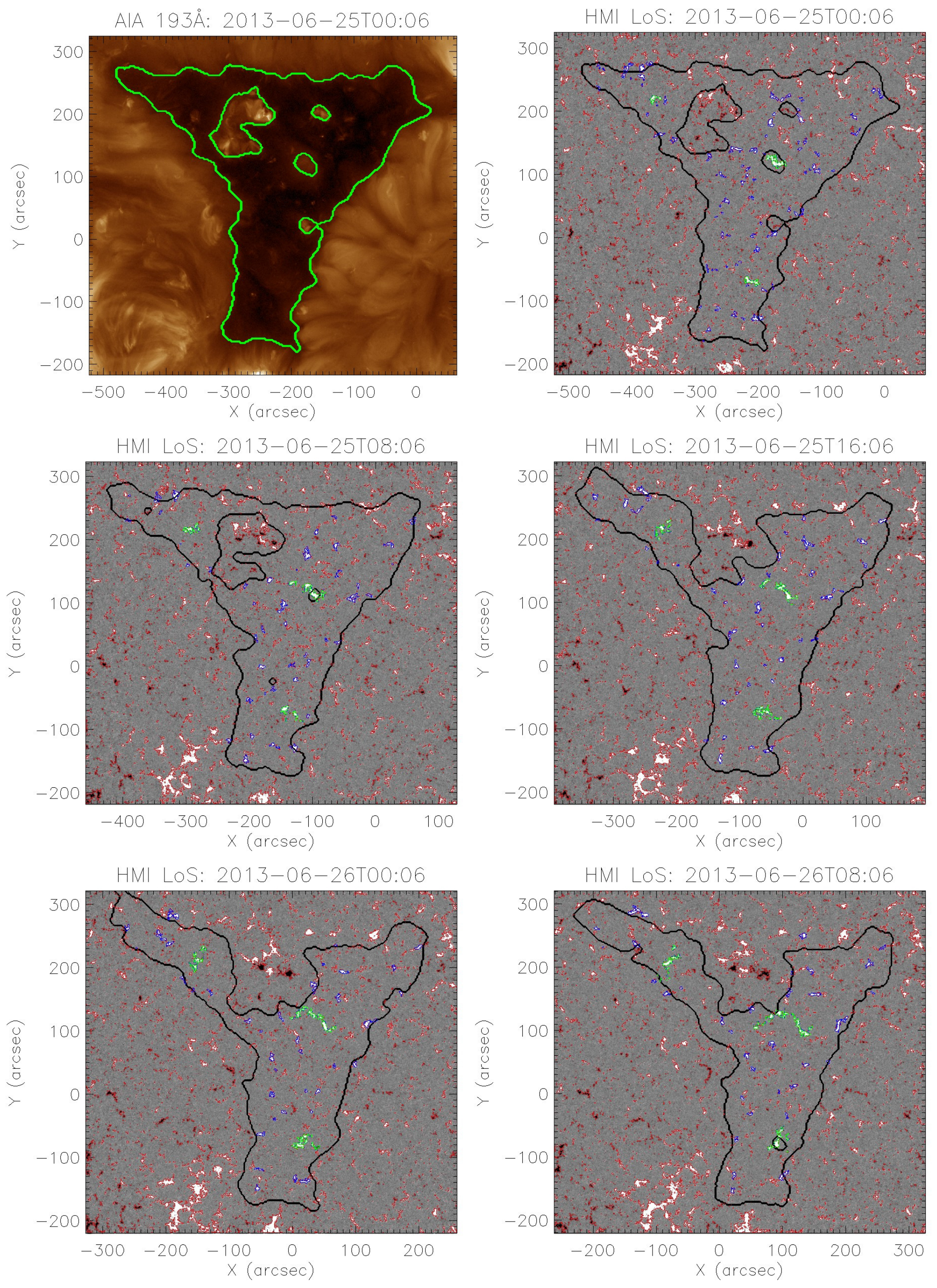}}

\caption{SDO/AIA \SI{193}{\angstrom} filtergram and HMI-LoS magnetograms, showing snapshots of the evolution of the photospheric magnetic field below the coronal hole on July 25 and 26, 2013. The magnetogram is scaled to \SI{\pm 100}{G} and the magnetic elements are outlined in red, blue, and green (see text). In the online version, the evolution of the magnetic elements is shown as a movie.}
\label{movieft}
\end{figure*}

Next, we focus on the temporal evolution of the magnetic elements. In the online version, Figure \ref{movieft} contains a movie of the temporal evolution of the magnetic field over two days at a cadence of six minutes. In this movie, all magnetic elements exceeding an absolute magnetic field strength of \SI{25}{G} are outlined in red, blue, and green. The magnetic elements as outlined in red within the coronal hole are small- to medium-sized and have a typical lifetime on the order of hours. These magnetic elements outline the magnetic network, and slowly reconfigure their position due to local flows produced by new emerging supergranular cells. The magnetic elements outlined in blue are further characterized by unusual high lifetimes of \SI{>2}{days}, i.e., they exist over the complete time range of the movie. They are mostly larger than the usual magnetic elements as outlined in red and located at the intersections of the magnetic network lanes. In addition, we have delineated three strongly magnetized elements in green. Tracking these elements, we can clearly see that some of them are growing and decaying, some break apart into multiple elements each then evolving individually, and others merge with other nearby elements. The splitting and merging shows that magnetic elements should not be seen as rigid structures, but more as a temporarily assembly of common substructures that we cannot distinguish with the available SDO/HMI spatial resolution. 
 
\begin{figure*}[tp]
\centering
\includegraphics[width = \textwidth]{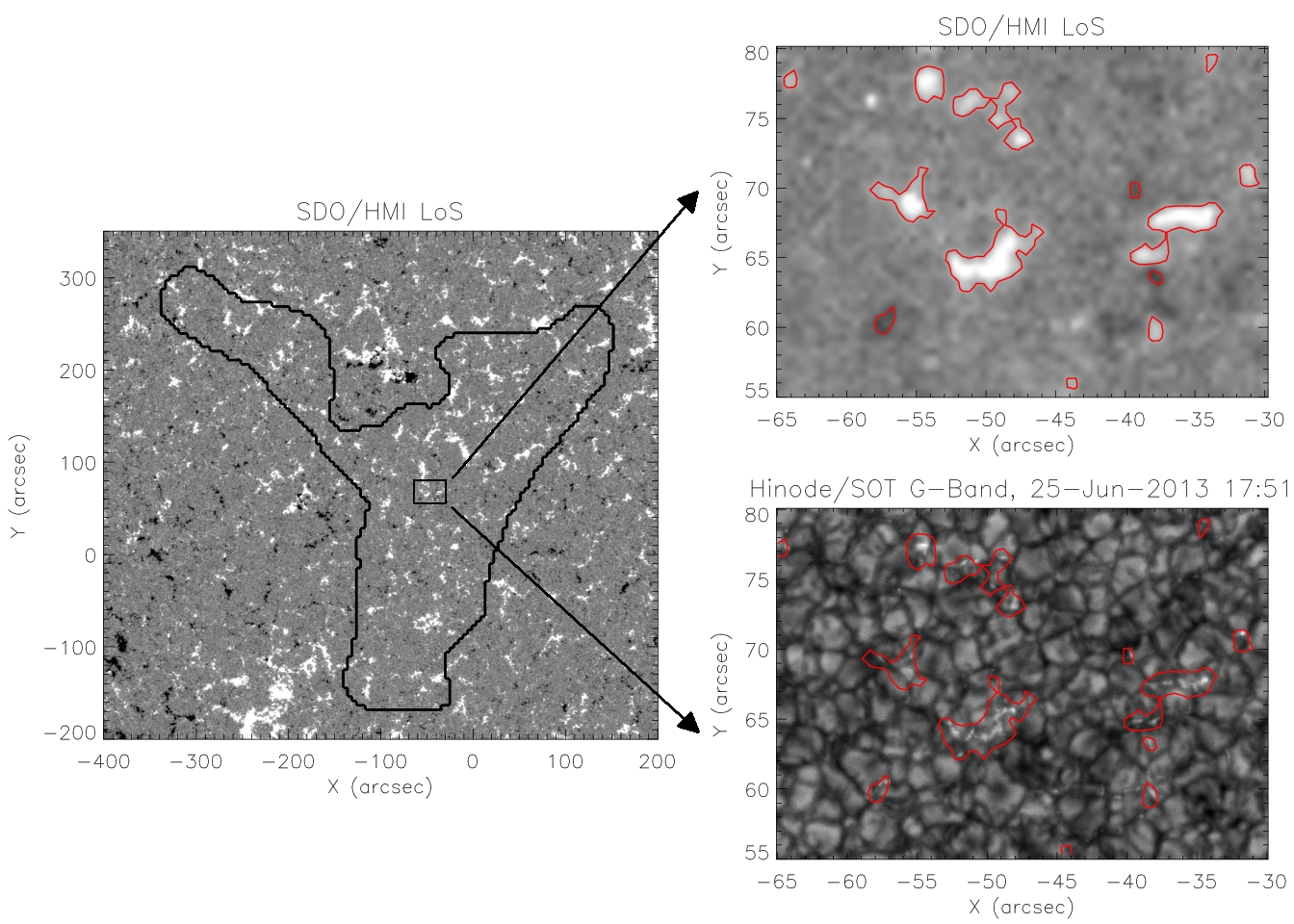}
\caption{SDO/HMI magnetogram of the coronal hole observed on July 25, 2013 (left). The magnetogram is scaled to \SI{\pm 25}{G}. The boundaries of the coronal hole as well as a small image section are outlined. The magnetogram section zoomed in (right top), and the corresponding Hinode/SOT G-band image of the image section (bottom right). The zoomed magnetogram is scaled to \SI{\pm 100}{G}. The contours of the magnetic elements as observed by SDO/HMI are overlayed in red. Within the magnetic elements, multiple photospheric MBPs are visible in the G-band image.}
\label{mbps}
\end{figure*}

When we look into the high-resolution G-band images of magnetic elements as observed by Hinode/SOT and overlay the contours of the magnetic elements as observed by SDO/HMI (Fig. \ref{mbps}), we find that many of the magnetic elements, and especially the large ones, contain more than one MBP in the G band. These MBPs are located at the intergranular lanes, and are known to be footprints of small scale expanding vertical magnetic fibers with magnetic field strength of \SIrange{1.2}{1.5}{kG} piercing through the photosphere \citep{utz2013}. Therefore, the magnetic elements as observed by SDO/HMI are clusters of vertical small-scale magnetic fibers, which cannot be resolved by SDO/HMI. This also explains why magnetic elements can split and evolve individually.

Summarizing, the photospheric area below coronal holes is characterized by weak magnetic fields in which a number of magnetic elements are embedded. Most of the magnetic elements in coronal holes are not located right below the darkest coronal regions, but surround them. They outline the magnetic network, and they can have lifetimes of more than two days. At high-resolution G-band observations, MBPs appear as the fine structure of magnetic elements. These are known to be the footpoints of vertical small-scale \si{kG}-field magnetic fibers piercing through the photosphere \citep{utz2013}.

\section{Magnetic elements: Photospheric magnetic fine-structure of coronal holes} \label{sec_structure}

In this section, we analyze the properties and distribution of magnetic elements in coronal holes. We start with the general properties of magnetic elements that contain only one MBP; this is a low-resolution proxy of the properties of the magnetic fibers as their substructure. Then, we extend the study to general magnetic elements containing multiple MBPs. Further, we analyze the distribution of the lifetimes of magnetic elements, which leads us to four classes of magnetic elements: magnetic elements related to the granulation, mesogranulation, supergranulation, and long-lived elements. Finally, we determine the distribution of the average properties of the magnetic elements in coronal holes depending on their lifetimes and the mean magnetic field strength of the overall coronal holes.

\subsection{Properties of magnetic fibers}
\label{subsec_cluster_mbp}

The magnetic elements observed by SDO/HMI containing only one MBP as seen by Hinode/SOT can be seen as low-resolution proxys of the individual MBPs, and thus of the underlying magnetic fibers. We assume that these are the basic magnetic building blocks of the magnetic structure of coronal holes (cf. Sect. \ref{sect_visual}).
Naturally, our dataset is limited to magnetic elements for which Hinode/SOT images were available, which leads us to a set of \num{1043} magnetic elements containing only one MBP.

\begin{figure}[tp]
\centering
\includegraphics[width = \linewidth]{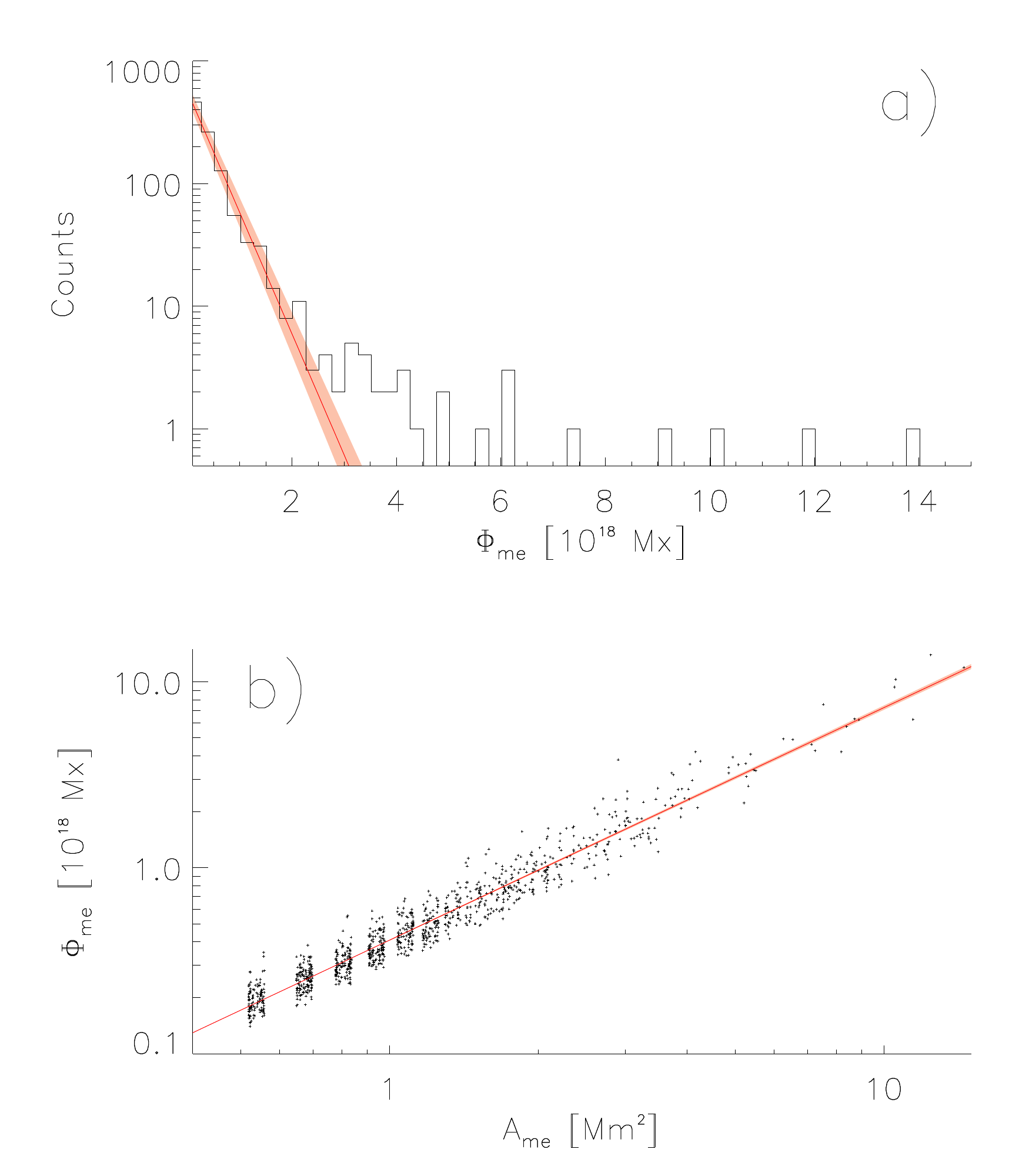}
\caption{Distribution of the magnetic fluxes of magnetic elements containing only one MBP (a). The corresponding fit only considers magnetic elements with \SI{<2e18}{Mx} (solid red line). Scatter plot of the magnetic fluxes of these magnetic elements vs. their areas and the corresponding fit (b). The shaded areas give the $1\sigma$ uncertainties of the corresponding fits.}
\label{magnetic_filament_1}
\end{figure}

In Figure \ref{magnetic_filament_1}a, we show the distribution of the magnetic flux arising from these magnetic elements. \SI{95}{\percent} of these magnetic elements have fluxes \SI{< 2e18}{Mx}, with the smallest flux of an element measured to be \SI{1.4e17}{Mx}. The distribution of the fluxes seem to follow an exponential law for fluxes  \SI{<2e18}{Mx}. However, to the right-hand tail of the distribution, the distribution starts to flatten and becomes more stable. Thus, an exponential fit would strongly underestimate the number of magnetic elements featuring larger magnetic fluxes. Since the number of magnetic elements with magnetic fluxes \SI{>2e18}{Mx} is relatively small, we acknowledge the possibility that at least a part of the stronger magnetic elements could contain multiple MBPs of which only one was identified by our extraction algorithm. Thus, we decided to use an exponential fit only for magnetic elements \SI{<2e18}{Mx}, and get for their number $N_{\text{me}}$, as function of magnetic flux $\Phi_\text{me}$,
\begin{equation}
N_\text{me} = (518 \pm 84) \cdot e^{ {-2.43 \pm 0.13} \cdot \left(\Phi_\text{me} / \SI{e18}{Mx}\right)} \label{eq_meexp}.
\end{equation}

In Figure \ref{magnetic_filament_1}b, we show the magnetic fluxes of these magnetic elements versus their areas in a double-logarithmic representation. We find a strong correlation between their fluxes and areas with a Spearman's correlation coefficient of $0.965 \pm 0.002$; the corresponding fit results in a power law as follows:  
\begin{equation}
\Phi_\text{me} = \SI{0.408 \pm 0.002e18}{Mx} \cdot \left( A_\text{me} / \si{Mm^{2}} \right)^{(1.251 \pm 0.009)}. \label{eq_flux_area_mc_single}
\end{equation}
Thus, the flux and area of magnetic elements containing only one MBP are strongly related to each other in a nonlinear way. The flux of a magnetic element is determined by its area, and with increasing area, its magnetic flux increases disproportionately faster.

\subsection{Properties of magnetic elements}
\label{subsec_magelement}

\begin{figure*}[tp]
\centering
\includegraphics[width = 1\textwidth]{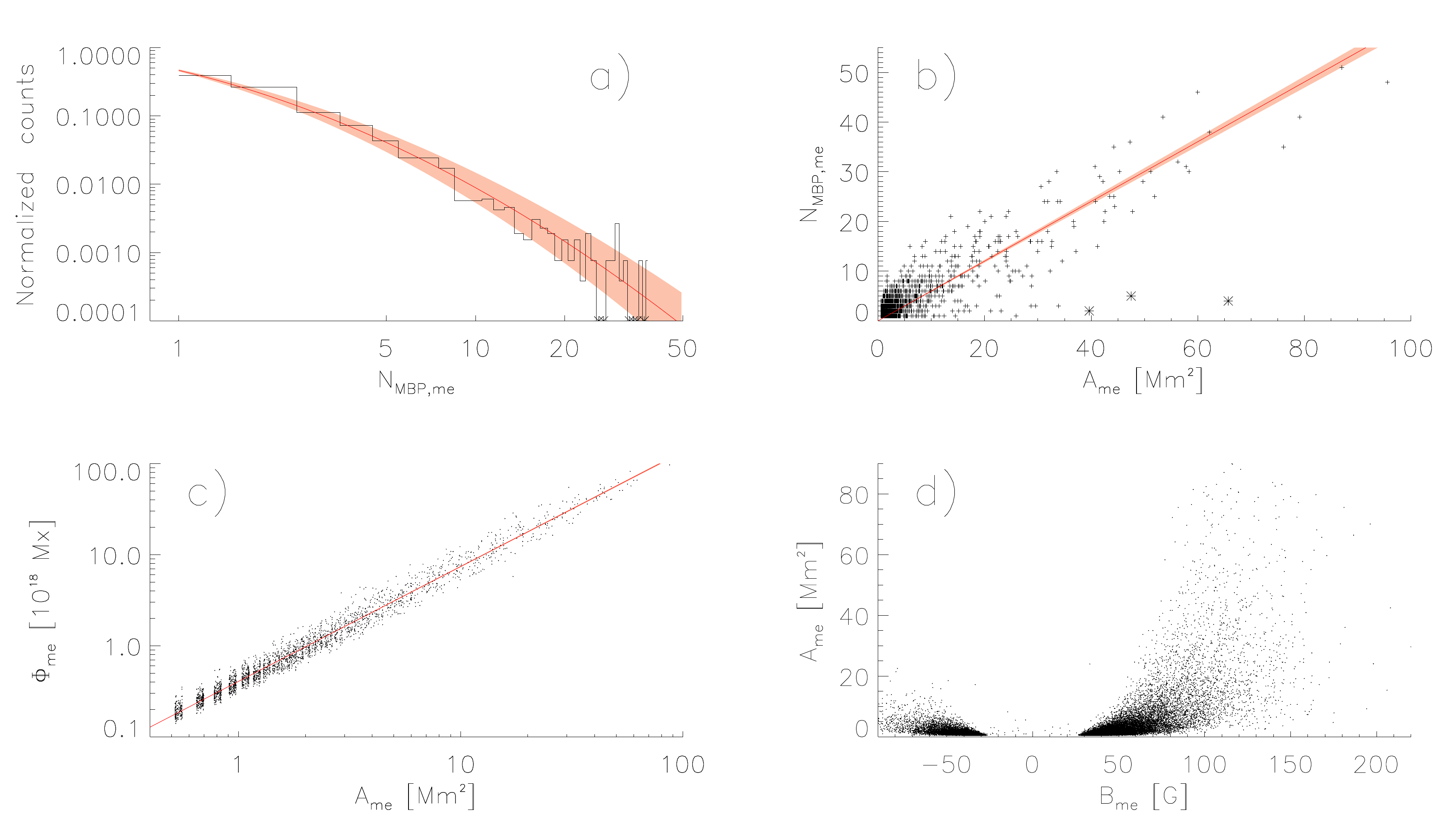}
\caption{Histogram of the number of MBPs per magnetic element; the asterisks denote unpopulated histogram bins (a). Number of MBPs per magnetic element vs. the area of the magnetic element; the asterisks denote the outliers (b). Magnetic flux of magnetic elements versus their areas (c). And areas of magnetic elements vs. their mean magnetic field densities (d). The shaded areas give the $1\sigma$ uncertainties of the corresponding fits.}
\label{magnetic_cluster_1}
\end{figure*}

In general, magnetic elements contain multiple magnetic fibers that are located so close to each other that they cannot be distinguished by HMI owing to the limited spatial resolution; they appear as one magnetic element.
By consulting the high-resolution Hinode/SOT data we can, however, get an estimate on the number of magnetic fibers per magnetic element by counting the number of observable MBPs. 
We note that in plots involving data obtained from MBPs, i.e., Figure \ref{magnetic_cluster_1}a and \ref{magnetic_cluster_1}b, the dataset is restricted to magnetic elements for which Hinode/SOT data was available; all other plots are derived from the complete dataset.

In Figure \ref{magnetic_cluster_1}a, we show the normalized histogram of the number of MBPs in a magnetic element. \SI{40}{\percent} of the magnetic elements have only one MBP and thus are likely individual magnetic elements, \SI{25}{\percent} contain two MBPs, and \SI{11}{\percent} three MBPs. Still \SI{4}{\percent} contain 5 bright points, and \SI{5}{\percent} more than 10 bright points; the largest element in our dataset contained even 51 bright points. Thus, it is very common that magnetic elements contain multiple MBPs. We derive the corresponding probability function to have N$_\text{MBP}$ MBPs in a magnetic element, $P (N_\text{MBP})$ by fitting the data with the constraint that the total probability of the probability function for having 1 to 51 MBPs is $1.00$, and get
\begin{equation}
\ln P = -\ (0.77 \pm 0.04)\ -\ (1.02 \pm 0.11) \cdot \ln N_\text{MBP}\ -\ (0.30 \pm 0.04) \cdot \ln^2 N_\text{MBP} . 
\end{equation}
The total probability of this probability function for having 1 to infinity MBPs is $1.001$ and thus only insignificantly larger than our constraint of $1.00$ for 1 to 51 MBPs. Under the assumption that our model is correct, this means that the probability of having a large number of MBPs in a magnetic element rapidly falls to zero, and that the probability function given is also valid for more general datasets.

The number of MBPs per magnetic elements versus the area of the magnetic element is shown in Figure \ref{magnetic_cluster_1}b. The three outliers delineated have been excluded from further analysis, since they contain multiple close-by MBPs that could not be distinguished by the segmentation algorithm. From the scatter plot, a clear correlation is apparent with a correlation coefficient of $cc = 0.83 \pm 0.01$. We fit the data under the constraint that a magnetic element of zero area has zero MBPs and get
\begin{equation}
N_\text{MBP,me} = (0.60 \pm 0.01) \cdot \left(A_\text{me} /\si{Mm^2} \right) . \label{eq_nmbp_area}
\end{equation}

The magnetic flux versus the area of the magnetic elements is plotted in Figure \ref{magnetic_cluster_1}c in a double-logarithmic representation. Again, a strong correlation at a Spearman's correlation coefficient of $0.984 \pm 0.001$ is visible, and the corresponding fit gives
\begin{equation}
\Phi_\text{me} = \SI{0.406 \pm 0.002e18}{Mx}  \cdot \left( A_\text{me} / \si{Mm^2} \right)^{(1.261 \pm 0.004)} . \label{eq_flux_area_mc}
\end{equation}
The magnetic elements containing multiple MBPs obey almost the same power law over three orders of magnitude as the magnetic elements containing only a single MBP (cf. Equ. \ref{eq_flux_area_mc_single}). Therefore, the flux of a magnetic element is in general determined by its area, and consecutively depends statistically on the number of its magnetic fibers (cf. Eq. \ref{eq_nmbp_area}).

Equation \ref{eq_flux_area_mc} has a further consequence: Since the flux of a magnetic element is increasing faster with increasing area, its mean magnetic flux density is also increasing, which is shown in Figure \ref{magnetic_cluster_1}d. The magnetic elements consisting of a small number of \num{< 5} MBPs only have an average mean magnetic field density of \SI{47 \pm 15}{G} at an average area of \SI{2 \pm 2.8}{Mm^2}. In contrast, the magnetic elements consisting of \numrange{40}{50} MBPs have an average mean magnetic field density of \SI{122 \pm 25}{G} at an average area of \SI{72 \pm 19}{Mm^2}.
Finally, we note that all the properties of the individual magnetic elements shown in this work are independent of the mean magnetic field strength of the corresponding coronal hole.

\subsection{Lifetimes of magnetic elements}
\label{sect_lifetime}

\begin{figure*}[tp]
\centering
\includegraphics[width = 1\textwidth]{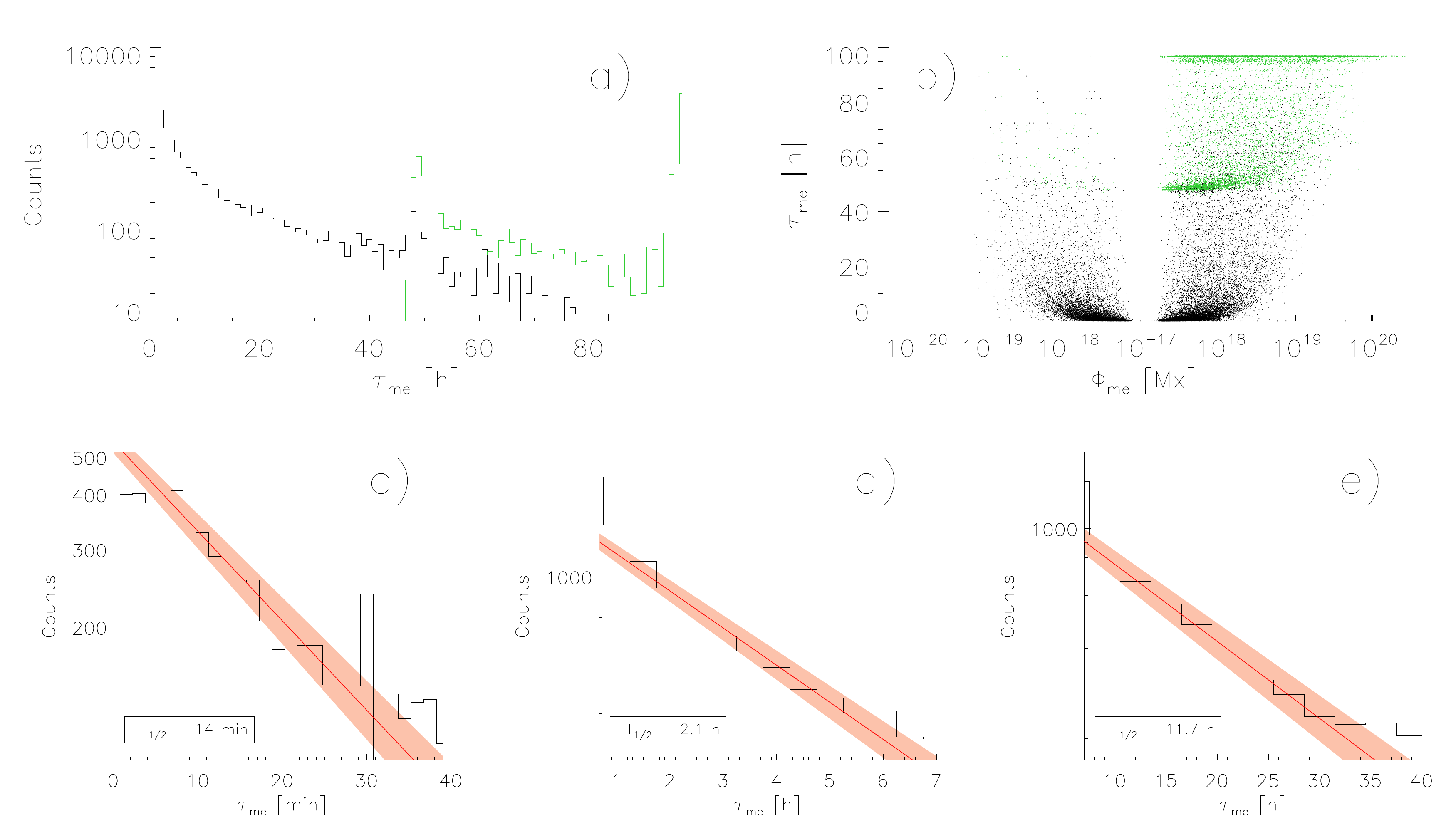}
\caption{Histogram of the lifetimes of magnetic elements (a), and zoom-ins (c, d, e). Lifetime of magnetic elements vs. their magnetic flux (b). The green line and green data points indicate minimum lifetimes related to magnetic elements that live longer than our tracking window of \SI{\pm 2}{days}. The red lines give the exponential fits and the shaded areas the corresponding $1\sigma$ uncertainties.}
\label{lifetimes}
\end{figure*}

Although the distributions of the areas and magnetic flux of all magnetic elements seem to be drawn from a single distribution, the lifetimes indicate that they are linked to different mechanisms.
Figure \ref{lifetimes}a shows the distribution of lifetimes of the magnetic elements; the green line indicates minimum lifetimes related to magnetic elements that live longer and therefore leave our tracking window of \SI{\pm 2}{days}. Of the magnetic elements \SI{21}{\percent} have lifetimes of less than \SI{40}{minutes}, \SI{29}{\percent} \SI{40}{minutes} to \SI{7}{hours}, \SI{17}{\percent} \SIrange{7}{40}{hours}, and \SI{33}{\percent} of more than \SI{40}{hours}. By zooming into the histogram (Fig. \ref{lifetimes}c-e), we find an exponential drop-off of the lifetimes in three time windows. Fitting the exponentially decreasing parts of the data in each of the windows, we find half-life times of \SI{14 \pm 1}{minutes} for the short-lived magnetic elements with lifetimes of \SIrange{5}{30}{minutes}, \SI{2.1 \pm 0.2}{hours} for the magnetic elements with lifetimes of \SIrange{1.5}{5.5}{hours}, and \SI{11.7 \pm 0.7}{hours} for magnetic elements with lifetimes of \SIrange{10}{30}{hours}. These half-life times match well the timescales of granulation, mesogranulation, and supergranulation. Therefore, we assign the magnetic elements into three corresponding categories dependent on their lifetime.

Further, by knowing the half-life time of the magnetic elements belonging to the class of supergranulation, i.e., \SI{11.7 \pm 0.7}{hours}, and their number at a given lifetime from Figure \ref{lifetimes}e, we can estimate how many magnetic elements with lifetimes of \SI{>4}{days} belong to the class of supergranulation by integrating the corresponding fit. Doing so, we find that only $26 \pm 9$ magnetic elements of the class of supergranulation should have lifetimes of \SI{>4}{days} in our dataset. However, by counting the magnetic elements with lifetimes \SI{>4}{days} detected in our dataset, we arrive at $3234$ magnetic elements. Therefore, the largest portion of the long-lived magnetic elements cannot belong to the class of supergranulation. They have to have another unknown origin giving us a fourth category of magnetic elements, the long-lived magnetic elements.

In Figure \ref{lifetimes}b, we plot the lifetimes of the magnetic elements versus their magnetic flux. Green data points again indicate minimum lifetimes related to magnetic elements that have left our tracking window, and therefore they naturally cluster at both borders of our tracking window, i.e., at \SI{48}{h} and \SI{96}{h}. We find that almost all of the short-lived magnetic elements with lifetimes up to \SI{20}{hours} appear with dominant and nondominant coronal hole polarity at similar rates. In contrast, \SI{96}{percent} of the long-lived magnetic elements with lifetimes \SI{>40}{hours} have the dominant coronal hole polarity, and they can contain significant more magnetic flux than the short-lived magnetic elements. All magnetic elements with a total magnetic flux of \SI{>2.5e19}{Mx} have long lifetimes of \SI{>40}{hours}, which  allows us to identify some of the long-lived magnetic elements just by magnetic flux measurements.

\subsection{Distribution of magnetic elements in coronal holes}

In the following, we analyze the distribution of magnetic elements in coronal holes dependent on their lifetimes and the mean magnetic field strength of the overall coronal hole $B_\text{CH}$. Thereby, in the subsequent plots, data corresponding to "weak" coronal holes, which we define as coronal holes with mean magnetic field densities of $|B_\text{CH}| < \SI{3}{G}$, are colored in blue, "medium" coronal holes with  $\SI{3}{G} < |B_\text{CH}| < \SI{5}{G}$ in red, and "strong" coronal holes with $|B_\text{CH}| > \SI{5}{G}$  in green.  
Further, we split the magnetic elements into four sets: all magnetic elements with lifetimes of \SI{<40}{minutes} are assigned to the set of  magnetic elements belonging to granulation (half-life time \SI{14}{minutes}), with lifetimes of \SI{40}{minutes} to \SI{7}{hours} to the set of mesogranulation (half-life time \SI{2.1}{hours}), with lifetimes of \SIrange{7}{40}{hours} to the set of supergranulation (half-life time \SI{11.7}{hours}), and with lifetimes \SI{>40}{hours} to the set of long-lived magnetic elements.

\subsubsection{Magnetic elements belonging to granulation, mesogranulation, and supergranulation}

\begin{figure*}[tp]
\centering
\includegraphics[width = 1.\textwidth]{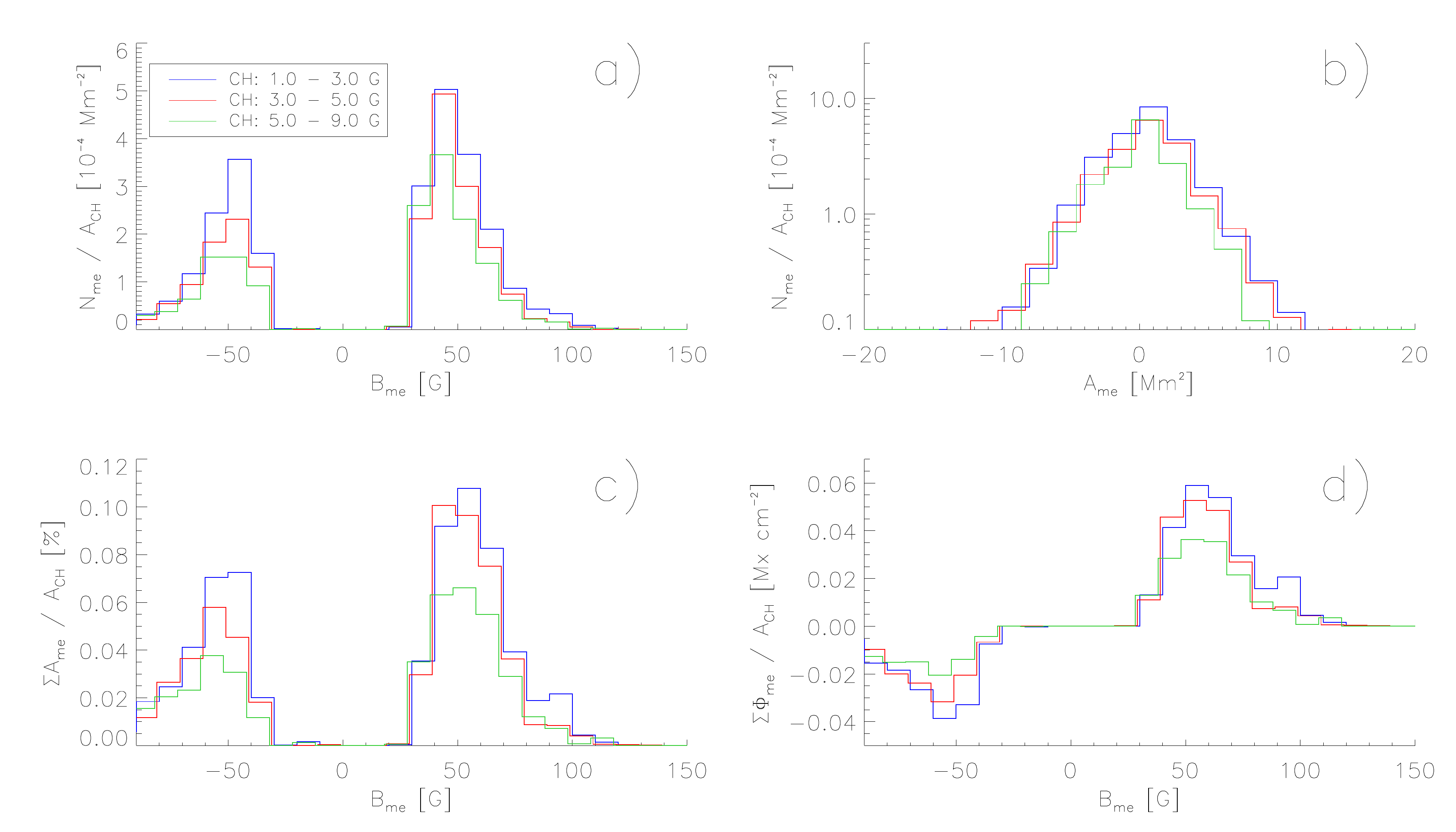}
\caption{Magnetic elements with lifetimes of \SIrange{7}{40}{hours}. Shown are the number density as a function of their mean magnetic flux densities (a) and areas (b), coverage of the coronal hole by the magnetic elements dependent on their mean magnetic flux density (c), and their contribution to the unbalanced magnetic flux of the overall coronal hole (d).}
\label{ft_small}
\end{figure*}

In Figure \ref{ft_small}, we show the distribution of the short-lived magnetic elements for the subset with lifetimes between \SIrange{7}{40}{hours}, i.e., magnetic elements belonging to the supergranulation.  
Figure \ref{ft_small}a and \ref{ft_small}b show the statistical number of the magnetic elements per coronal hole area dependent on the mean magnetic field density of the magnetic elements, respectively, on their area. We find a rather small number of magnetic elements belonging to the supergranulation, whereby the number of magnetic elements slightly decreases with increasing coronal hole strength: \SI{25 \pm 4}{\text{magnetic elements} / 10^4 Mm^2} for weak coronal holes,  \SI{21 \pm 5}{\text{magnetic elements} / 10^4 Mm^2} for medium coronal holes, and  \SI{16 \pm 4}{\text{magnetic elements} / 10^4 Mm^2} for strong coronal holes. Further, we find a slight imbalance in the distribution of polarities of the magnetic elements: \SI{36 \pm 10}{\percent} have the nondominant coronal hole polarity, and \SI{64 \pm 10}{\percent} the dominant coronal hole polarity. Almost all of the magnetic elements are small with an average area of \SI{2.7 \pm 2.2}{Mm^2}, and rather weak with an average mean magnetic field density of \SI{52 \pm 14}{G}.
Because of their rather small number and their small size, the magnetic elements belonging to the supergranulation cover only \SI{\approx 0.66 \pm 0.12}{\percent} of the coronal hole area, slightly dependent on the strength of the coronal hole (Fig. \ref{ft_small}c). Since the magnetic elements appear with both polarities at reasonable small rates and similar properties, their total unbalanced magnetic flux partially cancel each other (Fig. \ref{ft_small}d). 

\begin{figure*}[tp]
\centering
\includegraphics[width = 1\textwidth]{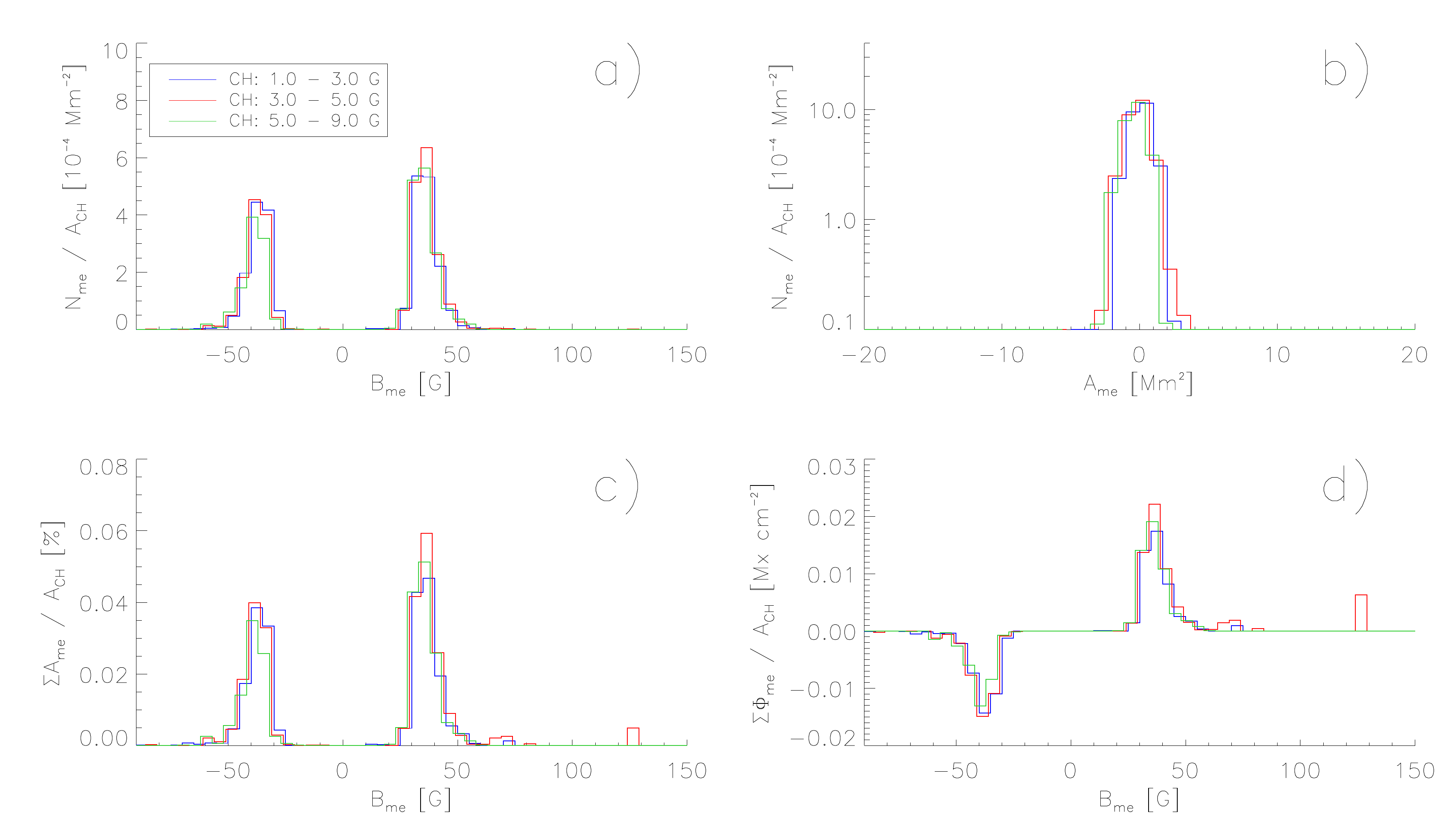}
\caption{As Figure \ref{ft_small}, but for magnetic elements with lifetimes \SI{<40}{minutes}.}
\label{ft_med}
\end{figure*}

\begin{figure*}[tp]
\centering
\includegraphics[width = 1\textwidth]{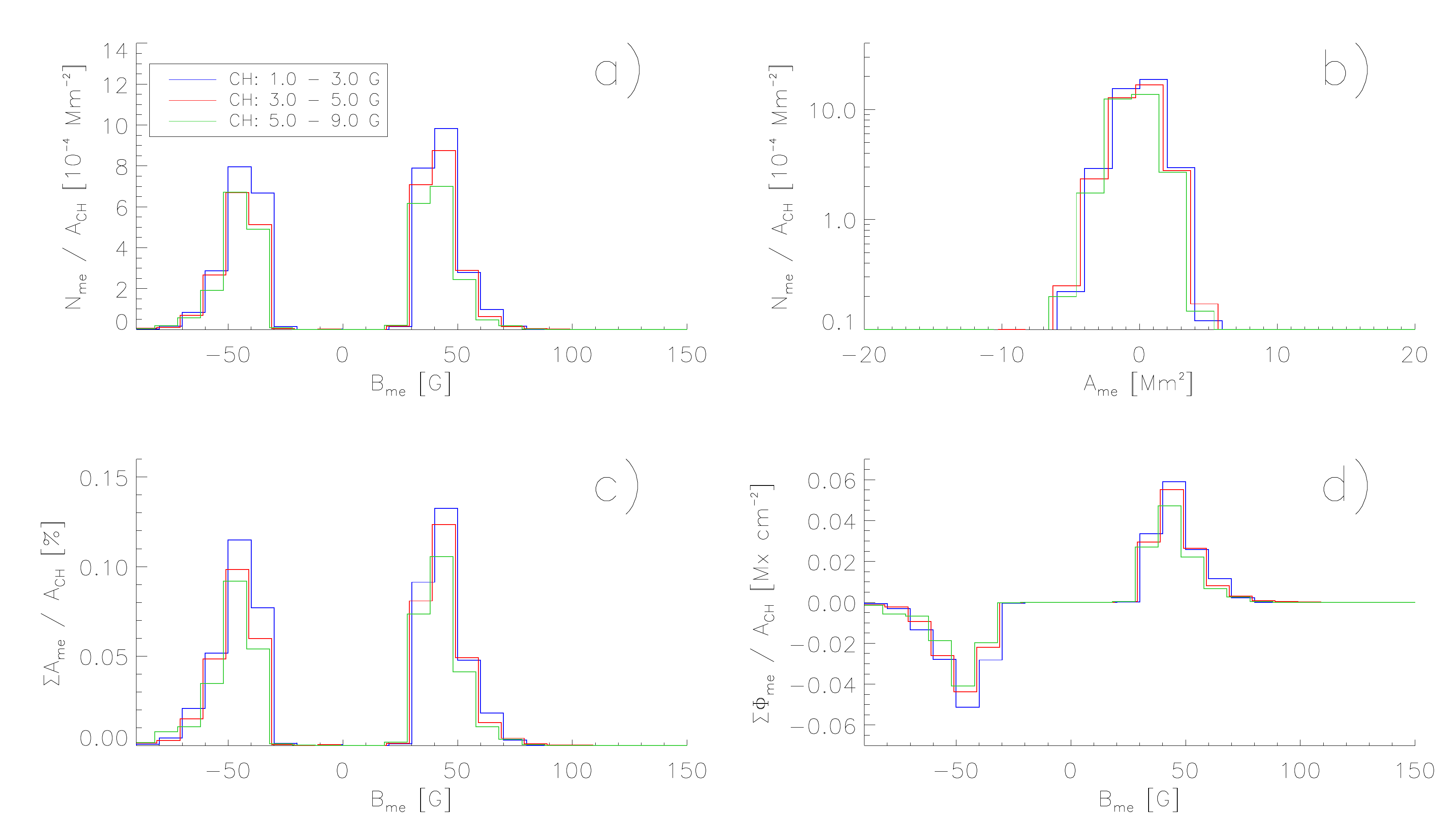}
\caption{As Figure \ref{ft_small}, but for magnetic elements with lifetimes of \SI{40}{minutes} to \SI{7}{hours}.}
\label{ft_large}
\end{figure*}

The corresponding distributions of magnetic elements belonging to granulation and mesogranulation are given in the Figures \ref{ft_med} and \ref{ft_large}. These distributions look in general similar to the distribution of magnetic elements belonging to the supergranulation, but with a much larger total number of magnetic elements and an even more balanced distribution of the polarities, and smaller areas and magnetic fluxes of the individual magnetic elements. The corresponding statistics are summarized in Table \ref{table_mc_stats}. 

\begin{figure}[tp]
\centering
\includegraphics[width = \linewidth]{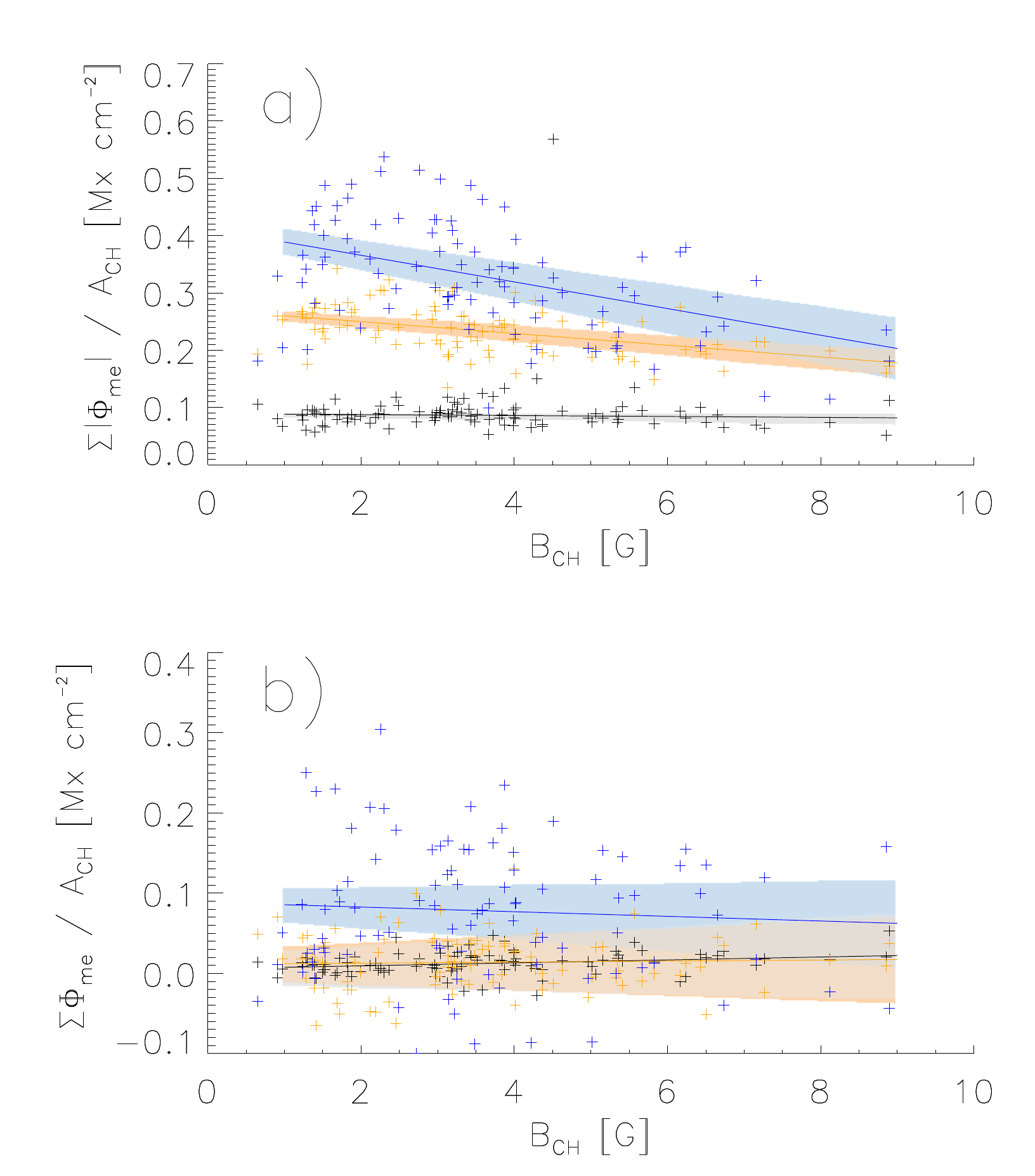}
\caption{Total unsigned (a) and total unbalanced (b) magnetic flux arising from magnetic elements with lifetimes of $<\SI{40}{minutes}$ (black), \SI{40}{minutes} to \SI{7}{hours} (orange), and \SIrange{7}{40}{hours} (blue) per coronal hole area vs. the absolute mean magnetic field densities of the coronal holes. The shaded areas give the $1\sigma$ uncertainties of the corresponding fits.}
\label{ft_small_integrated}
\end{figure}

By integrating over all magnetic elements which belong to granulation, mesogranulation, and supergranulation separately for each coronal hole, we find that the integrated unsigned magnetic flux of the magnetic elements per coronal hole area is dependent on the mean magnetic field strength of the overall coronal holes (Fig. \ref{ft_small_integrated}a). Coronal holes with a higher mean magnetic field strength have a significantly lower integrated unsigned magnetic flux per coronal hole area arising from these magnetic elements. The corresponding fits are given by
\begin{equation}
\sum |\Phi_\text{me}| / A_\text{CH} = 
\begin{cases}     
\SI{0.10 \pm 0.01}{Mx/cm^2} - (0.001 \pm 0.001)  \cdot B_\text{CH} ,\quad \text{for MEs belonging to granulation}, \\
\SI{0.28 \pm 0.01}{Mx/cm^2} - (0.012 \pm 0.002) \cdot B_\text{CH},\quad \text{for MEs belonging to mesogranulation}, \\
\SI{0.39 \pm 0.02}{Mx/cm^2} - (0.022 \pm 0.004) \cdot B_\text{CH},\quad \text{for MEs belonging to supergranulation}. \\
\end{cases} \label{udens_shortclusters}
\end{equation}

In contrast, their total unbalanced magnetic flux is only very weakly dependent on the mean magnetic field strength of the overall coronal hole. The magnetic elements belonging to granulation and mesogranulation contribute in total only \SI{0.02 \pm 0.05}{Mx/ cm^2} and \SI{0.03 \pm 0.02}{Mx/ cm^2}, respectively, to the unbalanced magnetic flux of the coronal holes, i.e., their unbalanced magnetic flux is negligible. The  magnetic elements belonging to the class of supergranulation contribute \SI{0.09 \pm 0.07}{Mx/ cm^2}; this flux density is still very small and maybe an artifact due to the overlapping lifetime distributions of magnetic elements belonging to the supergranulation and long-lived magnetic elements. 

In summary, the magnetic elements belonging to the granulation, mesogranulation, and supergranulation appear in total at a high rate with a rather balanced distribution between dominant and nondominant coronal hole polarities. These magnetic elements always have a small mean magnetic field density, small area, and only a small effective contribution to the unbalanced magnetic flux of the overall coronal hole. Because of these properties, we assume that these magnetic elements are mostly the footpoints of small-scale closed loops within coronal holes.

\begin{table*}[]
\resizebox{0.8\textwidth}{!}{
\begin{tabular}{lllll}
\multicolumn{1}{r}{Coronal holes:}                          & weak & medium & strong & all\\ 
\multicolumn{1}{r}{B$_\text{CH}$:}                          & \SIrange{1}{3}{G} & \SIrange{3}{5}{G} & \SI{>5}{G} & \multicolumn{1}{r}{\SI{>1}{G}} \\ \hline
\\
Magnetic elements belonging to                                 &                   &                   & \multicolumn{1}{r}{}           \\[.2cm]
\underline{Granulation, $\tau_\text{me} < $ \SI{40}{min}}                   &                   &                   & & \multicolumn{1}{r}{}           \\
$\sum \Phi_\text{me} / A_\text{CH}$ [\si{Mx/cm^2}]        & $0.01 \pm 0.01$  & $0.03 \pm 0.07$  &  $0.02 \pm 0.01$  & $0.02 \pm 0.05$      \\
$\sum |\Phi_\text{me}| / A_\text{CH}$ [\si{Mx/cm^2}]      & $0.08 \pm 0.01$  & $0.10 \pm 0.08$  &  $0.08 \pm 0.02$  & $0.09 \pm 0.05$      \\
$N_\text{me} / A_\text{CH}$ [\SI{1/e4}{Mm^2}]                   & $27 \pm 4$       & $28 \pm 5$       &  $25 \pm 4$       & $27 \pm 5$                       \\
Percent with dominant CH polarity [\%]                         & $55 \pm 5$      & $58 \pm 7$        &  $61 \pm 7$       & $58 \pm 7$                 \\
$\sum A_\text{me} / A_\text{CH}$ [\%]                          & $0.23 \pm 0.03$  & $0.26 \pm 0.11$  & $0.22 \pm 0.04$   & $0.24 \pm 0.07$                 \\
$\sum \Phi_\text{me} / \Phi_\text{CH}$  [\%]                   & $0.5 \pm 0.4$   & $0.8 \pm 0.2$    & $0.3 \pm 0.2$      & $0.6 \pm 0.01$                  \\
$\sum |\Phi_\text{me}| / \Phi_\text{CH,us}$ [\%]               & $0.9 \pm 0.1$   & $1.0 \pm 0.2$     & $0.6 \pm 0.1$     & $0.8 \pm 0.5$                      \\
                                    &                            &                   &                   &                                \\
\underline{Mesogranulation,  $\tau_\text{me}$ = \SI{40}{min} to \SI{7}{h}} &                   &                   &               &                  \\
$\sum \Phi_\text{me} / A_\text{CH}$ [\si{Mx/cm^2}]        & $0.03 \pm 0.02$   & $0.03 \pm 0.02$   & $0.02 \pm 0.02$  & $0.03 \pm 0.02$                       \\
$\sum |\Phi_\text{me}| / A_\text{CH}$ [\si{Mx/cm^2}]      & $0.34 \pm 0.03$   & $0.23 \pm 0.04$   & $0.20 \pm 0.03$  & $0.23 \pm 0.04$                         \\
$N_\text{me} / A_\text{CH}$ [\SI{1/e4}{Mm^2}]                   & $41 \pm 6$        & $35 \pm 7$        & $31 \pm 5$       & $36 \pm 7$                       \\
Percent with dominant CH polarity [\%]                         & $54 \pm 6$        & $56 \pm 7$        & $53 \pm 7$       & $55 \pm 7$                      \\
$\sum A_\text{me} / A_\text{CH}$ [\%]                          & $0.56 \pm 0.07$   & $0.50 \pm 0.09$    & $0.43 \pm 0.06$ & $0.51 \pm 0.09$                       \\
$\sum \Phi_\text{me} / \Phi_\text{CH}$ [\%]                    & $1.8 \pm 1.1$     & $0.7 \pm 0.7$     & $0.4 \pm 0.3$    & $1.0 \pm 1.0$                       \\
$\sum |\Phi_\text{me}| / \Phi_\text{CH,us}$ [\%]               & $2.7 \pm 0.4$     & $2.1 \pm 0.4$     & $1.5 \pm 0.3$    & $2.2 \pm 0.6$                     \\
                              &                                  &                   &                   &                                \\
\underline{Supergranulation, $\tau_\text{me} = $ \SIrange{7}{40}{h}}        &                   &                   &                         &        \\
$\sum \Phi_\text{me} / A_\text{CH}$ [\si{Mx/cm^2}]        & $0.10 \pm 0.08$    & $0.09 \pm 0.06$   & $0.08 \pm 0.05$  & $0.09 \pm 0.07$                     \\
$\sum |\Phi_\text{me}| / A_\text{CH}$ [\si{Mx/cm^2}]      & $0.38 \pm 0.08$    & $0.33 \pm 0.03$   & $0.24 \pm 0.08$  & $0.33 \pm 0.11$                     \\
$N_\text{me} / A_\text{CH}$ [\SI{1/e4}{Mm^2}]                   & $25 \pm 4$         & $21 \pm 5$        & $16 \pm 4$       & $21 \pm 6$                \\
Percent with dominant CH polarity [\%]                         & $61 \pm 9$         & $64 \pm 10$       & $68 \pm 11$      & $64 \pm 10$                       \\
$\sum A_\text{me} / A_\text{CH}$ [\%]                          & $0.66 \pm 0.12$    & $0.57 \pm 0.17$    & $0.42 \pm 0.11$ & $0.57 \pm 0.17$                    \\
$\sum \Phi_\text{me} / \Phi_\text{CH}$ [\%]                    & $5.3 \pm 4.7$      & $2.6 \pm 1.7$     & $1.4 \pm 1.0$    & $3.3 \pm 3.4$                     \\
$\sum |\Phi_\text{me}| / \Phi_\text{CH,us}$ [\%]               & $3.9 \pm 0.8$      & $3.0 \pm 1.0$     & $1.9 \pm 0.6$    & $3.1 \pm 1.1$                        \\        \\    

\underline{Long-lived magnetic elements $\tau_\text{me} > $ \SI{40}{h}}   &                   &                   &                        &        \\
$\sum \Phi_\text{me} / A_\text{CH}$ [\si{Mx/cm^2}]        & $1.3 \pm 0.5$    & $2.6 \pm 0.5$    & $4.8 \pm 1.0$  &  $2.6 \pm 1.5$                         \\
$\sum |\Phi_\text{me}| / A_\text{CH}$ [\si{Mx/cm^2}]      & $1.3 \pm 0.5$    & $2.7 \pm 0.5$    & $4.9 \pm 1.0$  & $2.7 \pm 1.5$                        \\
$N_\text{me} / A_\text{CH}$ [\SI{1/e4}{Mm^2}]                   & $29 \pm 8$       & $44 \pm 7$        & $62 \pm 9$      & $42 \pm 14$                         \\
Percent with dominant CH polarity [\%]                         & $94 \pm 4$       & $97 \pm 3$        & $99 \pm 1$     & $96 \pm 4$                        \\
$\sum A_\text{me} / A_\text{CH}$ [\%]                          & $1.7 \pm 0.5$    & $3.1 \pm 0.5$     & $5.1 \pm 0.9$  & $3.0 \pm 1.4$                         \\
$\sum \Phi_\text{me} / \Phi_\text{CH}$ [\%]                    & $63 \pm 8$       & $71 \pm 6$        & $76 \pm 3$     & $69 \pm 8$                       \\
$\sum |\Phi_\text{me}| / \Phi_\text{CH,us}$ [\%]               & $14 \pm 4$       & $24 \pm 4$        & $37 \pm 5$     & $23 \pm 10$                      \\
                              &                                 &                   &                   &                                \\
\underline{Quiet CH regions}               &                                &                   &                   &                                \\[.2cm]
$\Phi_\text{qu} / A_\text{CH}$ [\si{Mx/cm^2}]             & $0.49 \pm 0.35$   & $0.80 \pm 0.11$    & $1.2 \pm 0.22$  & $0.77 \pm 0.29$                     \\
$\Phi_\text{qu,us} / A_\text{CH}$ [\si{Mx/cm^2}]          & $7.1 \pm 0.3$     & $7.1 \pm 0.3$     & $7.1 \pm 0.2$    & $7.1 \pm 0.3$                     \\
$A_\text{qu} / A_\text{CH}$ [\%]                               & $95 \pm 0.7$      & $93 \pm 0.7$      & $92 \pm 1.0$     & $93 \pm 1.4$                     \\
$\Phi_\text{qu} / \Phi_\text{CH}$ [\%]                         & $26 \pm 4$        & $22 \pm 3$        & $18 \pm 2$       & $22 \pm 4$   \\
$\Phi_\text{qu,us}/ \Phi_\text{CH,us}$ [\%]                    & $73 \pm 4$        & $64 \pm 2$        & $54 \pm 3$       & $65 \pm 8$
\end{tabular}}

\caption{Source of the magnetic flux of coronal holes. The table lists the properties of the magnetic elements belonging to granulation, mesogranulation, supergranulation, long-lived magnetic elements, and the properties of the quiet coronal hole regions dependent on the mean magnetic field densities of the coronal holes. Each subsection gives the unbalanced and total unsigned magnetic flux arising from these regions, the average number of magnetic elements per area, the percentage of how many magnetic elements have the dominant coronal hole polarity, the coverage of the coronal hole in percent, and their contribution to the unbalanced and total unsigned magnetic flux of the coronal hole. We note that \SI{\approx 4}{\percent} of the total unbalanced magnetic flux $\Phi_\text{CH}$ and \SI{\approx 6}{\percent} of the total unsigned magnetic flux $\Phi_\text{CH,us}$ arises from isolated pixels with flux densities \SI{>25}{G,} which are not included in any sections of the table.}
\label{table_mc_stats}  
\end{table*}

\subsubsection{Long-lived magnetic elements}

In this section, we investigate the properties of the long-lived magnetic elements with lifetimes of \SI{> 40}{hours}. These correspond to the large magnetic elements of Figure \ref{movieft} denoted in blue and green, and according to Figure \ref{lifetimes}b they are expected to strongly contribute to the unbalanced magnetic flux of coronal holes.
 
\begin{figure*}[tp]
\centering
\includegraphics[width = 1\textwidth]{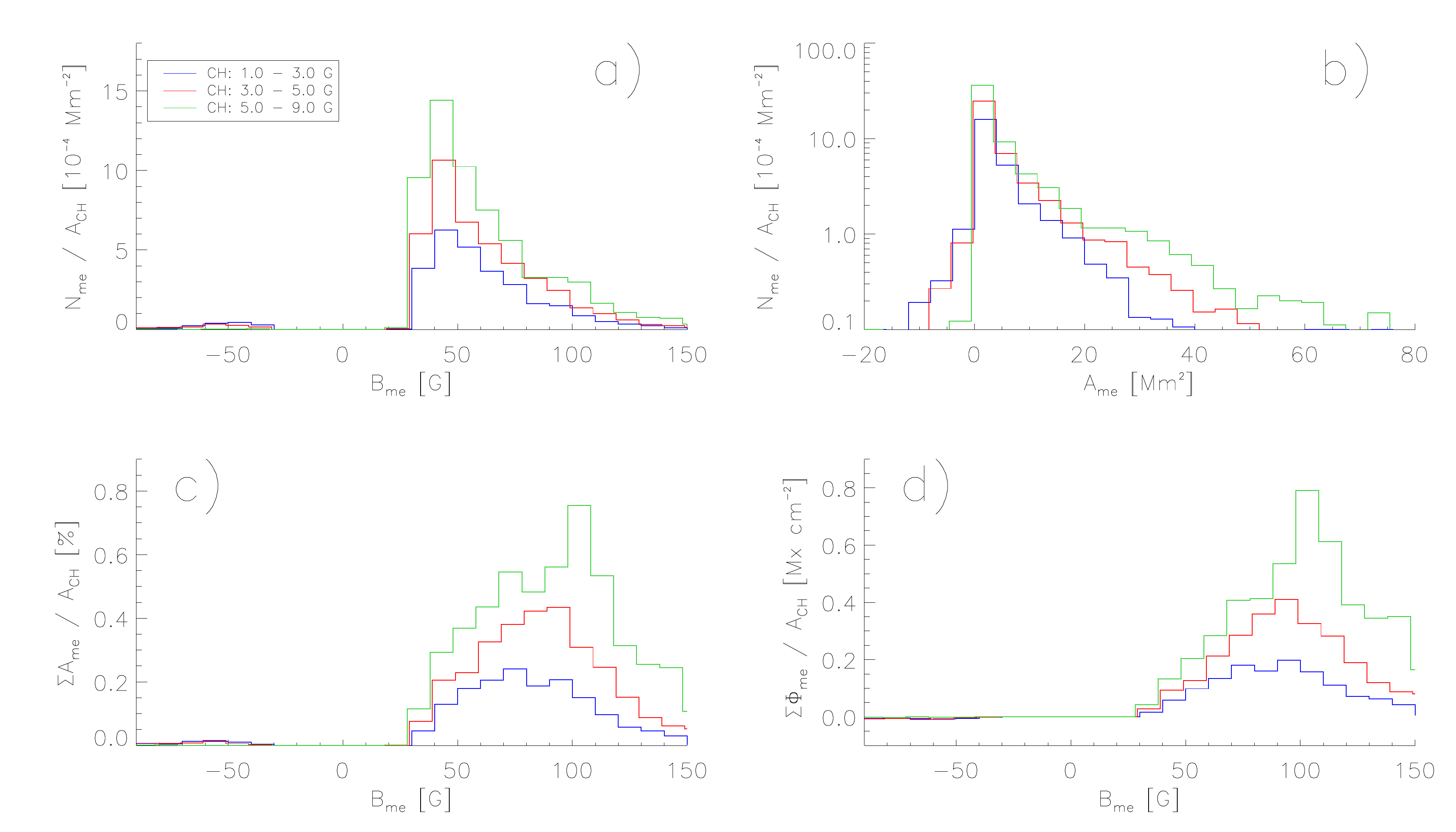}
\caption{As Figure \ref{ft_small}, but for magnetic elements with lifetimes \SI{>40}{hours}.}
\label{ft_long1}
\end{figure*}

In Figure \ref{ft_long1}a and \ref{ft_long1}b, we show the statistical number of long-lived magnetic elements per coronal hole area, dependent on their mean magnetic field density and area. We find a strong asymmetry in the number of magnetic elements of dominant and nondominant polarity. The number of magnetic elements with the nondominant coronal hole polarity is very small and decreases with increasing coronal hole strength: \SI{1.7  \pm 1.1}{\text{magnetic elements} / 10^4 Mm^2} for weak coronal holes,  \SI{1.5 \pm 1.0}{\text{magnetic elements} / 10^4 Mm^2} for medium coronal holes, and  \SI{0.69 \pm 0.38}{\text{magnetic elements} / 10^4 Mm^2} for strong coronal holes. In contrast, the number of magnetic elements with the dominant coronal hole polarity strongly increases with the strength of the coronal hole; we find \SI{27 \pm 8}{\text{magnetic elements} / 10^4 Mm^2} for weak coronal holes,  \SI{43 \pm 7}{\text{magnetic elements} / 10^4 Mm^2} for medium coronal holes, and  \SI{63 \pm 9}{\text{magnetic elements} / 10^4 Mm^2} for strong coronal holes. Most of the magnetic elements have rather small magnetic field densities showing an average of \SI{63 \pm 25}{G} and average areas of \SI{7.3 \pm 12}{Mm^2}; however, we note that a small number reach values of up to \SI{150}{G} and \SI{80}{Mm^2}.
This means that there are a small number of magnetic elements with a large mean magnetic field density, a large area, and thus possessing a large magnetic flux.

In Figure \ref{ft_long1}c, we show the total area that long-lived magnetic elements cover within the coronal hole. Clearly, the coverage depends on the strength of the coronal hole, with a total coverage of \SI{1.7 \pm 0.5}{\percent} for weak coronal holes, \SI{3.1 \pm 0.5}{\percent} for medium coronal holes, and \SI{5.1 \pm 0.9}{\percent} for large coronal holes. Thereby again, the long-lived magnetic elements with nondominant polarity have a negligible contribution to the coverage of \SI{< 0.05}{\percent}.
It is important to note that in Figure \ref{ft_long1}c the peak of the coverages changes with the strength of the coronal hole: whereas for weak coronal holes magnetic elements at \SI{70}{G} cover the largest area, for strong coronal holes magnetic elements in the range of \SI{100}{G} contribute the most to the area. 

In Figure \ref{ft_long1}d, we show the contribution of the long-lived magnetic elements to the unbalanced magnetic flux. Apparently, the long-lived magnetic elements with nondominant polarity are negligible and they only contribute a magnetic flux density of \SI{0.031 \pm 0.037}{Mx / cm^2} to the nondominant polarity. In contrast, the magnetic flux arising from magnetic elements with dominant coronal hole polarity is much larger and dependent on the strength of the coronal hole:  on the average \SI{1.3 \pm 0.5}{Mx /cm^2} arise from these magnetic elements in weak coronal holes of \SIrange{1}{3}{G}, \SI{2.7 \pm 0.5}{Mx / cm^2}  for medium coronal holes of \SIrange{3}{5}{G}, and \SI{5.3 \pm 1.8}{Mx / cm^2}  for strong coronal holes of \SI{>5}{G}. This confirms that a large portion of the unbalanced magnetic flux of coronal holes arises from these magnetic elements. Further, we note from Figure \ref{ft_long1}d that the peak contribution to the unbalanced magnetic flux, and especially the peak height, changes with the strength of the coronal holes: whereas for weak coronal holes the peak contribution is \SI{0.19}{Mx / cm^2} arising from magnetic elements of \SIrange{90}{100}{G}, for strong coronal holes the peak contribution is  \SI{0.75}{Mx / cm^2} arising from magnetic elements in the range \SIrange{100}{110}{G}. 

\begin{figure}[tp]
\centering
\includegraphics[width = \linewidth]{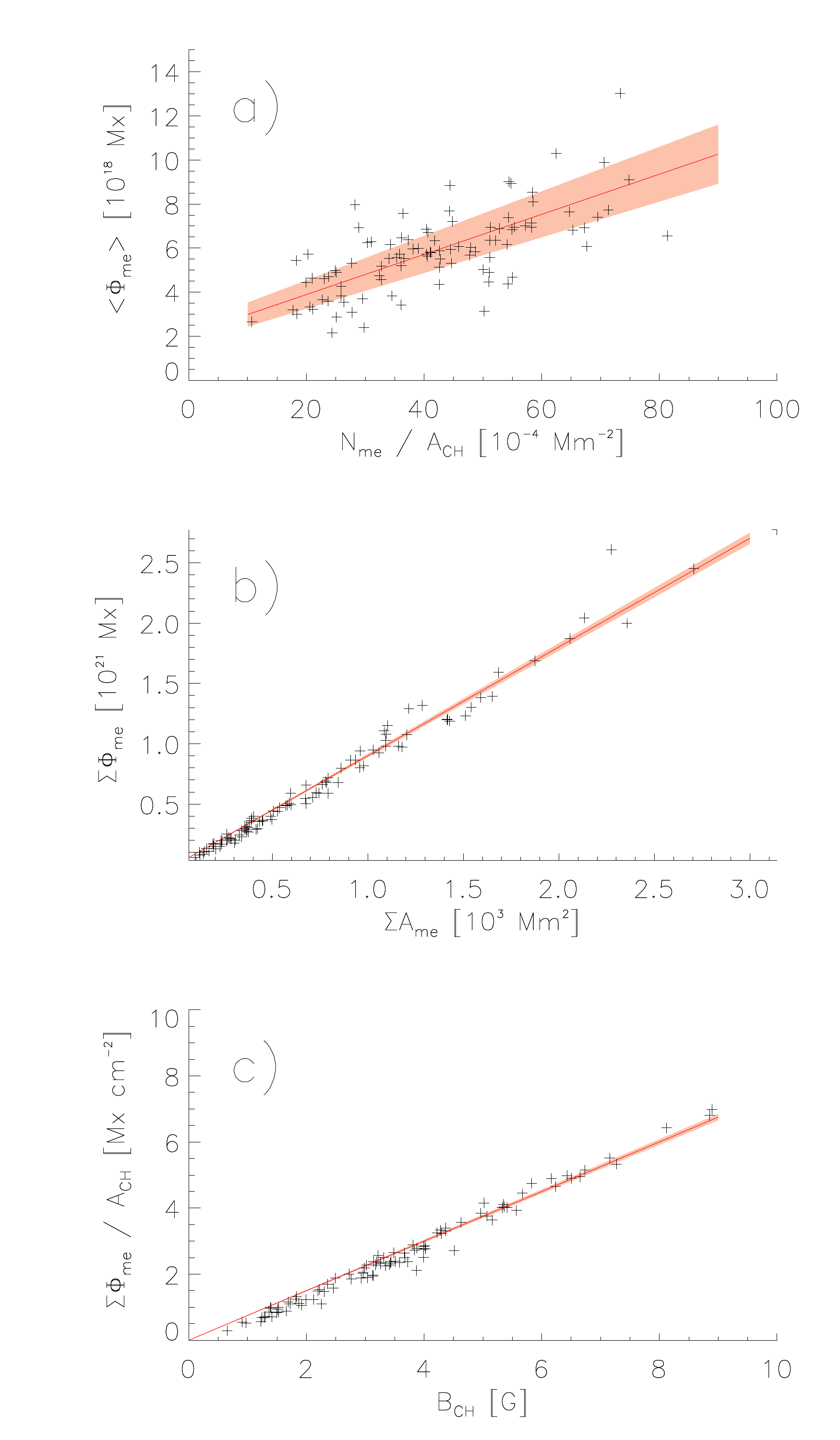}
\caption{Density of magnetic elements with lifetimes of \SI{>40}{hours} vs. their average magnetic flux derived for each coronal hole (a). Total unbalanced magnetic flux which arises from these magnetic elements versus the total area they cover (b). Total unbalanced magnetic flux that arises from theses magnetic elements per coronal hole area vs. the absolute mean magnetic field densities of the coronal holes (c). The shaded areas give the $1\sigma$ uncertainties of the corresponding fits.}
\label{ft_large_integrated}
\end{figure}

Finally, we integrate over the properties of the long-lived magnetic elements for each coronal hole. In Figure \ref{ft_large_integrated}a, we show the average number of magnetic elements per coronal hole area versus their averaged magnetic flux. We find a reasonable correlation with a correlation coefficient of $0.68 \pm 0.05$, the corresponding fit is
\begin{equation}
 <\Phi_\text{me}> = \SI{2.07 \pm 0.09e18}{Mx} + \SI{9.1 \pm 1.0e20}{Mx Mm^2} \cdot  N_\text{me} / A_\text{CH} . \label{eq_nmc_flux}
\end{equation}
This means that in coronal holes, the number of long-lived magnetic elements and their average individual flux are related to each other. 

Figure \ref{ft_large_integrated}b shows the unbalanced magnetic flux that arises from the long-lived magnetic elements versus the total area they cover. Clearly, their total unbalanced magnetic flux is strongly dependent on the area they cover with a correlation coefficient of $0.99 \pm 0.004$. The fit under the constraint that zero area results in zero flux is given by 
\begin{equation}
\sum_\text{me} \Phi_\text{me} =  \SI{0.90 \pm 0.02e18}{Mx} \cdot \sum_\text{me} A_\text{me} /\si{Mm^2}.
\end{equation}
This relationship can be explained by magnetic fibers as the substructure of magnetic elements. The total area the long-lived magnetic elements cover within a coronal hole is a measure of the total number of magnetic fibers they contain; and since the flux of the magnetic fibers is statistically distributed, this area is also a measure of their total magnetic flux (cf. Sect. \ref{subsec_cluster_mbp} and \ref{subsec_magelement}). Further, note that according to Figure \ref{ft_long1}a and \ref{ft_long1}c most of the total area they cover stems from a small number of large magnetic elements, meaning that most of the magnetic fibers and thus of the total magnetic flux are concentrated in these.

In Figure \ref{ft_large_integrated}c, we show the total unbalanced magnetic flux that arises from the long-lived magnetic elements of a coronal hole per coronal hole area versus the mean magnetic field strengths of the coronal holes. We find a clear linear relation between the magnetic flux arising from these magnetic elements and the mean magnetic field strengths of the coronal holes with a correlation coefficient of $cc=0.994 \pm 0.0018$. At least \SI{44}{\percent}, on average \SI{69 \pm 8}{\percent}, and maximal \SI{84}{\percent} of the mean magnetic field strength of the overall coronal hole has its origin in the long-lived magnetic elements. The linear fit under the assumption that the unbalanced magnetic flux of the long-lived magnetic elements goes to zero when the mean magnetic field density of the coronal holes goes to zero gives 
\begin{equation}
\sum_\text{me} \Phi_\text{me} / A_\text{CH} = (0.75 \pm 0.01) \cdot B_\text{CH}, 
\end{equation}
and follows reasonably well the distribution of data points. According to the fit, we slightly underestimate the measured magnetic flux arising from the long-lived magnetic elements for weak coronal holes. Therefore, the derived contributions of the long-lived magnetic elements to the unbalanced magnetic flux of the overall coronal hole gives a lower boundary.

In summary, we find that the total flux of the long-lived magnetic elements with nondominant polarity is almost negligible. The number of magnetic elements with dominant polarity as well as their strength are strongly increasing with increasing mean magnetic field strength of the coronal hole. The total area these magnetic elements cover determines their total magnetic flux, and the largest contribution to the magnetic flux of the coronal hole arises especially from a small number of large, strong magnetic elements. On the average, the magnetic elements contain \SI{\approx 69}{\percent} of the unbalanced magnetic flux of the overall coronal hole. Thus, we conclude that these are the footpoints of the magnetic funnels that mostly determine the open magnetic topology of the coronal hole.

\section{Distribution of the magnetic field density in quiet coronal hole regions: Unresolved magnetic elements or a background magnetic field}
\label{sec_quietch}

In this section, we analyze the magnetic field distribution of the quiet coronal hole regions, i.e., the regions without magnetic elements. On average, the quiet coronal hole regions cover \SI{93 \pm 1.4}{\percent} of the overall coronal hole and contain \SI{22 \pm 4}{\percent} of the unbalanced magnetic flux of the coronal hole. Basically, the unbalanced magnetic flux arising from the quiet coronal hole regions can have its origin in (1) an omnipresent small background magnetic field, which should lead to a slight shift of the whole magnetic field strength distribution to the dominant coronal hole polarity. However, if we have such an offset, it would also be strongly blurred out by the noise in the magnetograms which makes it hard to detect. Or (2), the unbalanced magnetic flux could have its origin in unresolved magnetic elements. These are isolated distinct features with higher mean magnetic field densities and thus should be found on the top end of the magnetic field distribution of quiet coronal hole regions. 

\begin{figure}[tp]
\centering
\includegraphics[width = \linewidth]{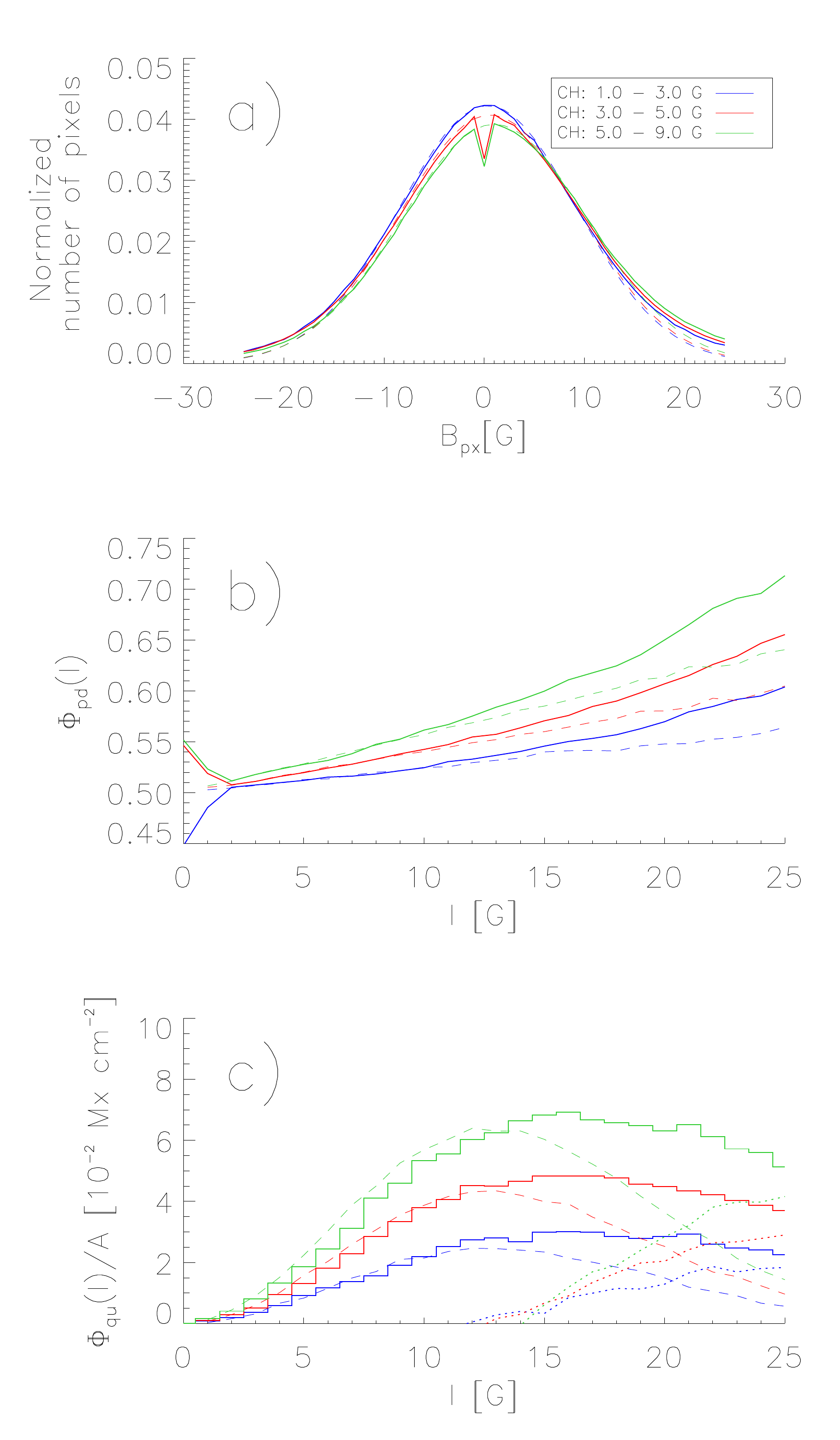}
\caption{Average distribution of the magnetic field densities in coronal holes (a). Imbalance of the magnetic field at a given magnetic level $l$ (b), and total unbalanced magnetic flux arising from pixels at a magnetic level $l$ (c). The solid lines are the measured distributions and the dashed lines correspond to the fitted background magnetic field. In panel (c), the dotted lines give the remaining unbalanced magnetic flux of the quiet coronal hole regions, which is not explained by a background magnetic field. }
\label{quiet_dominance}
\end{figure}

We start by examining the possibility of a small background magnetic field within coronal holes. Hereby, we model the lower end of the magnetic field strength distribution as a superposition of an omnipresent background magnetic field and noise on the data. 
First, we determine the average noise level in the magnetograms within the coronal holes from the magnetic field distribution (Fig. \ref{quiet_dominance}a). Although Poisson noise is the main physical reason for the noise in the underlying HMI filtergrams, the noise in the final magnetogram data products is determined by additive noise \citep{deForest2017}. In order to determine the additive noise level, we fit a Gaussian distribution to the core of the magnetic field distribution. Thereby, we use linearly decreasing weights with increasing distance from the Gaussian center to allow slight deviations at the wings due to possibly unresolved magnetic elements. This fit results in an average additive noise level of \SI{9.2}{G}. 

Next, we investigate the strength of the background magnetic field. To do so, in Figure \ref{quiet_dominance}b, we plot the polarity dominance within the quiet coronal hole regions versus the magnetic field level $l$ within the coronal holes, dependent on the strength of coronal holes. Hereby, we assign all pixels with $l - \SI{1}{G} < |B_{px}| < l + \SI{1}{G}$ to the magnetic field level $l$, and define the polarity dominance at a given magnetic field level $l$  as the ratio of the magnetic flux arising from these pixels with dominant coronal hole polarity to the magnetic flux arising from these pixels with both polarities, $\Phi_\text{pd}(l) = \Phi(l) / \left( \Phi(l) + \Phi(-l)\right)$. At a symmetric magnetic field strength distribution, the polarity dominance would be $\Phi_\text{pd}(l) = 0.5$ for all magnetic field levels $l$, whereas for an unipolar magnetic field strength distribution it would be $\phi_\text{pd}(l) = 1$ for all $l$. Therefore, the polarity dominance is a measure of the asymmetry of the magnetic field strength distribution with respect to the point of origin evaluated at a given magnetic field level $l$. Therefore, it is in particular a very sensitive measure on slight shifts of a symmetric additive noise distribution toward the dominant polarity. 
In order to determine whether a slight shift of the measured magnetic field distribution toward the dominant polarity is existent, we first create an artificial magnetic field strength distribution consisting of a slight uniform background magnetic field superposed with additive noise at level of \SI{9.2}{G}. Then, we vary the strength of the background magnetic field until the polarity dominance of the artificial magnetic field strength distribution fits the polarity dominance of the measured magnetic field strength distribution for small magnetic field strength \SI{<9}{G}, i.e., for the core of the magnetic field strength distribution. This guarantees that especially the core of the magnetic field strength distribution, which consists of a superposition of a possible small background magnetic field blurred by strong noise, is well fit.
Applying this method to the total magnetic field distribution consisting of all weak, medium, and strong coronal holes, we find an average background magnetic field of \SI{0.49 \pm 0.35}{G} for weak coronal holes, \SI{0.80 \pm 0.11}{G} for medium coronal holes, and \SI{1.2 \pm 0.22}{G} for strong coronal holes. 

Next, we investigate how much the background magnetic field contributes to the unbalanced magnetic flux of the quiet coronal hole regions. We plot the total unbalanced magnetic flux of the overall quiet coronal hole regions that arises from a magnetic level $l$ within the coronal holes versus the magnetic level $l$ (Fig. \ref{quiet_dominance}c, solid lines). The first impression is that most of the unbalanced magnetic flux of the quiet coronal hole regions arises from rather high magnetic levels $l > \SI{10}{G}$. However, when we again create an artificial magnetic field distribution resembling our noise level and the background magnetic field and compare its unbalanced magnetic flux with the unbalanced flux of the overall quiet coronal hole regions (dashed lines), we find that large parts of the measured distribution of the unbalanced magnetic flux can be explained by the background magnetic field superposed with additive noise. By having the distribution of the unbalanced magnetic flux arising from the blurred background magnetic field, we can derive the missing unbalanced magnetic flux that is needed to end up at the total measured unbalanced magnetic flux distribution (dotted lines). This distribution is clearly smaller than that of the blurred background magnetic field. Further, it is located at the top end of the magnetic field distribution, relating it to magnetic elements that were not resolved by our extraction algorithm.

Next, we determine the functional relationship of the unbalanced magnetic flux that arises from the background magnetic field and unresolved magnetic elements, respectively, on the strength of the individual coronal holes. Thereby, we apply the procedure described above to each coronal hole individually: First, we fit its noise level and strength of the background magnetic field; we then use these values to create an artificial magnetic field distribution resembling the background magnetic field superposed with noise, and subtract this artificial background magnetic field distribution from the real magnetic field distribution of the coronal hole to obtain the magnetic field distribution of its presumingly unresolved magnetic elements. Finally, we derive the unbalanced magnetic flux arising from the background field and from its unresolved magnetic elements separately by integrating over the corresponding derived magnetic field distributions.

\begin{figure*}[tp]
\centering
\includegraphics[width = 1.\textwidth]{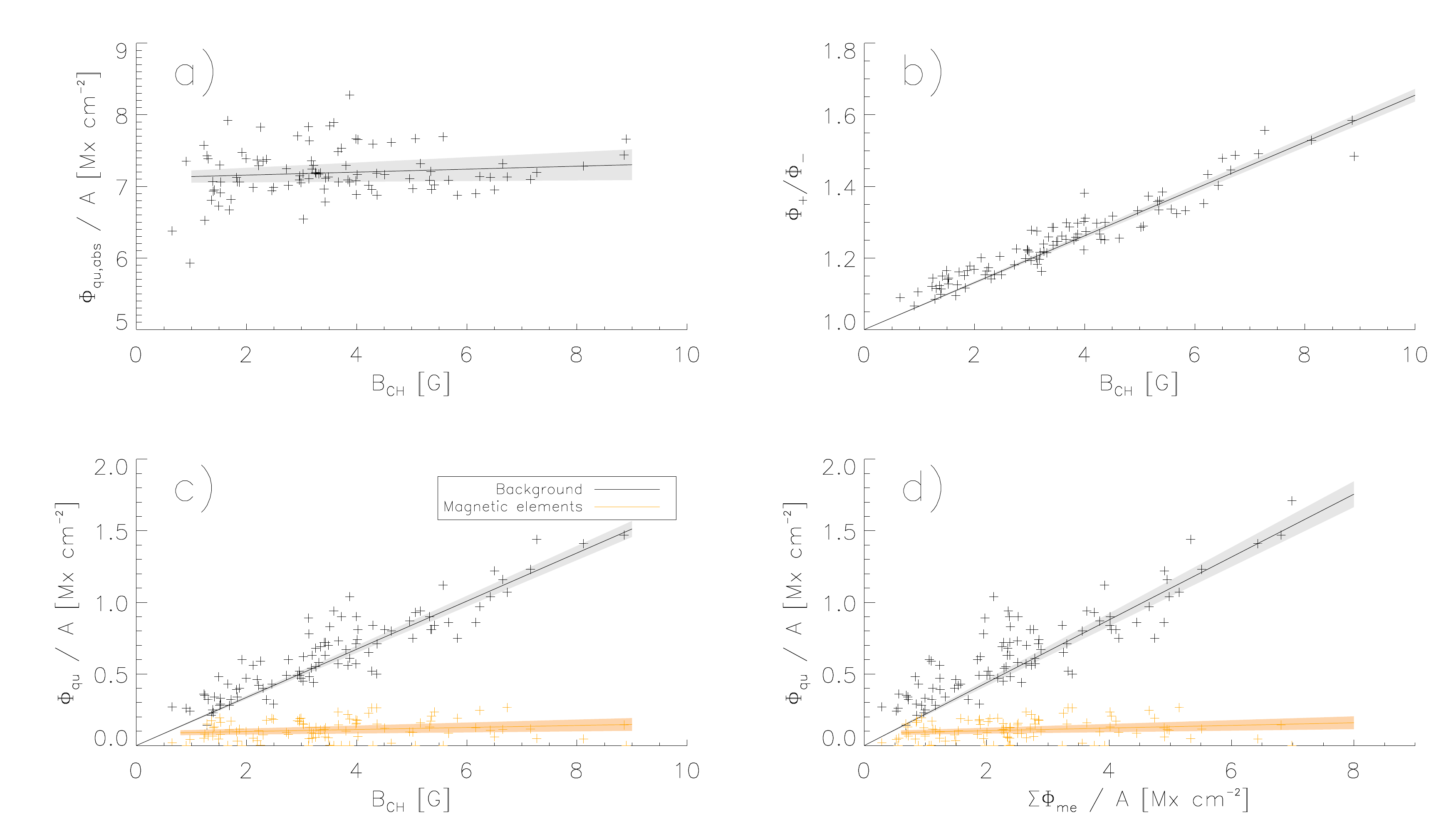}
\caption{Total unsigned magnetic flux of the quiet coronal hole regions per area (a), and magnetic flux imbalance of quiet coronal hole regions (b) vs. the mean magnetic field strength of the overall coronal holes. Unbalanced magnetic flux per area ,which arises from quiet coronal hole regions vs. the mean magnetic field density of the coronal holes (c). Unbalanced magnetic flux per area, which arises from quiet coronal hole regions vs. the unbalanced magnetic flux per area, which arises from the long-lived magnetic elements (d). In panel (c) and (d), the unbalanced magnetic flux arising from the quiet coronal hole regions was split into the magnetic flux, which arises from a background magnetic field (black) and from unresolved magnetic elements (orange). The shaded areas give the $1\sigma$ uncertainties of the corresponding fits.}
\label{quiet_fluxtube}
\end{figure*}

In Figure \ref{quiet_fluxtube}a, we start with plotting the total unsigned magnetic flux per coronal hole area, which arises from the overall quiet coronal hole regions versus the mean magnetic field strength of the coronal holes. We do not find a dependence on the mean magnetic field strength of the coronal hole; the total unsigned flux per area is on average constant at about \SI{7.1 \pm 0.3}{Mx/cm^2},  indicating that the total unsigned magnetic flux in the quiet coronal hole regions  is dominated by magnetogram noise.

In Figure \ref{quiet_fluxtube}b, we show the imbalance of the magnetic flux arising from quiet coronal hole regions of dominant to nondominant polarity. The imbalance is clearly linearly increasing with increasing mean magnetic field strength of the overall coronal hole with a Pearson's correlation coefficient of $cc = 0.95 \pm 0.01$. Fitting the data under the constraint that an overall mean magnetic field strength of \SI{0}{G} is equatable with having a balanced magnetic flux, we get
\begin{equation}
\Phi_+ / \Phi_- = 1.0 + (0.065 \pm 0.002) \cdot B_\text{CH} / \si{G}, 
\end{equation}

Correspondingly, we check how this affects the unbalanced magnetic flux arising from the quiet coronal hole regions (Fig. \ref{quiet_fluxtube}c). The unbalanced magnetic flux per coronal hole area due to the background magnetic field (black data points) is linearly dependent on the mean magnetic field strength of the coronal hole with a correlation coefficient of $0.91 \pm 0.02$. Demanding that the background magnetic field turns to zero for very weak coronal holes of \SI{\approx 0}{G}, the fit gives 
\begin{equation}
\Phi_\text{qs} / A_\text{CH} =  (0.17 \pm 0.006) \cdot B_\text{CH}. 
\end{equation}
According to the fit, we slightly overestimate the background magnetic flux density determined from the data, which does not however have strong effects on our results. 
In contrast, the unbalanced magnetic flux per area arising from the unresolved magnetic elements (yellow data points) is very weak and almost independent on the mean magnetic field strength of the coronal holes, similar to the magnetic elements belonging to granulation and mesogranulation. Their contribution to the unbalanced magnetic flux is on average constant at \SI{0.11 \pm 0.07}{Mx/cm^2}.

Finally, in Figure \ref{quiet_fluxtube}d, we plot the unbalanced magnetic flux arising from the quiet coronal hole regions versus the unbalanced magnetic flux arising from the long-lived magnetic elements, both normalized to the coronal hole areas. The magnetic flux arising from the background magnetic field is strongly correlated to the magnetic flux arising from the long-lived magnetic elements with a correlation coefficient of $0.88 \pm 0.02$. Under the constraint that a weak coronal hole of \SI{\approx 0}{G} should have neither a background magnetic field nor unbalanced magnetic flux from the long-lived magnetic elements, the fit to the data gives
\begin{equation}
\Phi_\text{qu,bg} / A_\text{CH} =(0.22 \pm 0.01) \cdot \sum_\text{me} \Phi_\text{me} / A_\text{CH} . \label{cor_bg_long}
\end{equation}
Therefore, with increasing flux from the long-lived magnetic elements the unbalanced magnetic flux from the background magnetic field also increases.
The magnetic flux arising from the unresolved magnetic elements is again independent of the magnetic flux arising from the long-lived magnetic elements.

Summarizing this section, we find that most of the unbalanced magnetic flux arising from the quiet coronal hole regions has its source in a background magnetic field dependent on the mean magnetic field strength of the overall coronal hole. Further, this background magnetic field is well correlated to the magnetic flux arising from the long-lived magnetic elements. The magnetic flux arising from the unresolved magnetic elements in the quiet coronal hole regions is rather small and independent of the strength of the coronal holes, similar to the magnetic flux arising from magnetic elements belonging to granulation, mesogranulation, and supergranulation.

\section{Overall magnetic properties of coronal holes as a result of their inner structure} \label{sec_overall}

Until now, we have determined the distribution of the magnetic elements and of the quiet coronal hole regions depending on the strength of the coronal holes. But in fact, we swapped the cause and effect; their properties determine the magnetic characteristics of the overall coronal hole. In this section, we derive and explain the main magnetic parameters of coronal holes, i.e., their mean unsigned magnetic field strength, mean magnetic field strength, unbalanced magnetic flux, and percentaged unbalanced magnetic flux, by their inner magnetic fine structure. 

According to the previous sections, the structure of coronal holes can be divided into a background magnetic field, short- and medium-lived magnetic elements belonging to the granulation, mesogranulation, and supergranulation, and long-lived magnetic elements. The background magnetic field is strongly blurred by noise and thus gives a considerable contribution to the mean unsigned magnetic field strength and a small to medium contribution to the unbalanced magnetic flux. The short- and medium-lived magnetic elements are mostly well balanced and cover in total only about \SI{1.3 \pm 0.24}{\percent} of the coronal hole, thus they only give a very small contribution to the mean unsigned magnetic field strength and unbalanced magnetic flux. The long-lived magnetic elements have almost all the same polarity; they determine the magnetic field polarity of the coronal hole and give the strongest contribution to the unbalanced magnetic flux. Further, the magnetic elements contain fine-structure, magnetic fibers that can be observed as photospheric MBPs. Since the magnetic flux of the magnetic fibers is statistically distributed, the total flux of the magnetic fibers in the overall coronal hole should be given by their number, or correspondingly by the total area they cover. The total area the magnetic fibers cover, on the other hand, is given by the total area the magnetic elements cover.

\begin{figure}[tp]
\centering
\includegraphics[width = \linewidth]{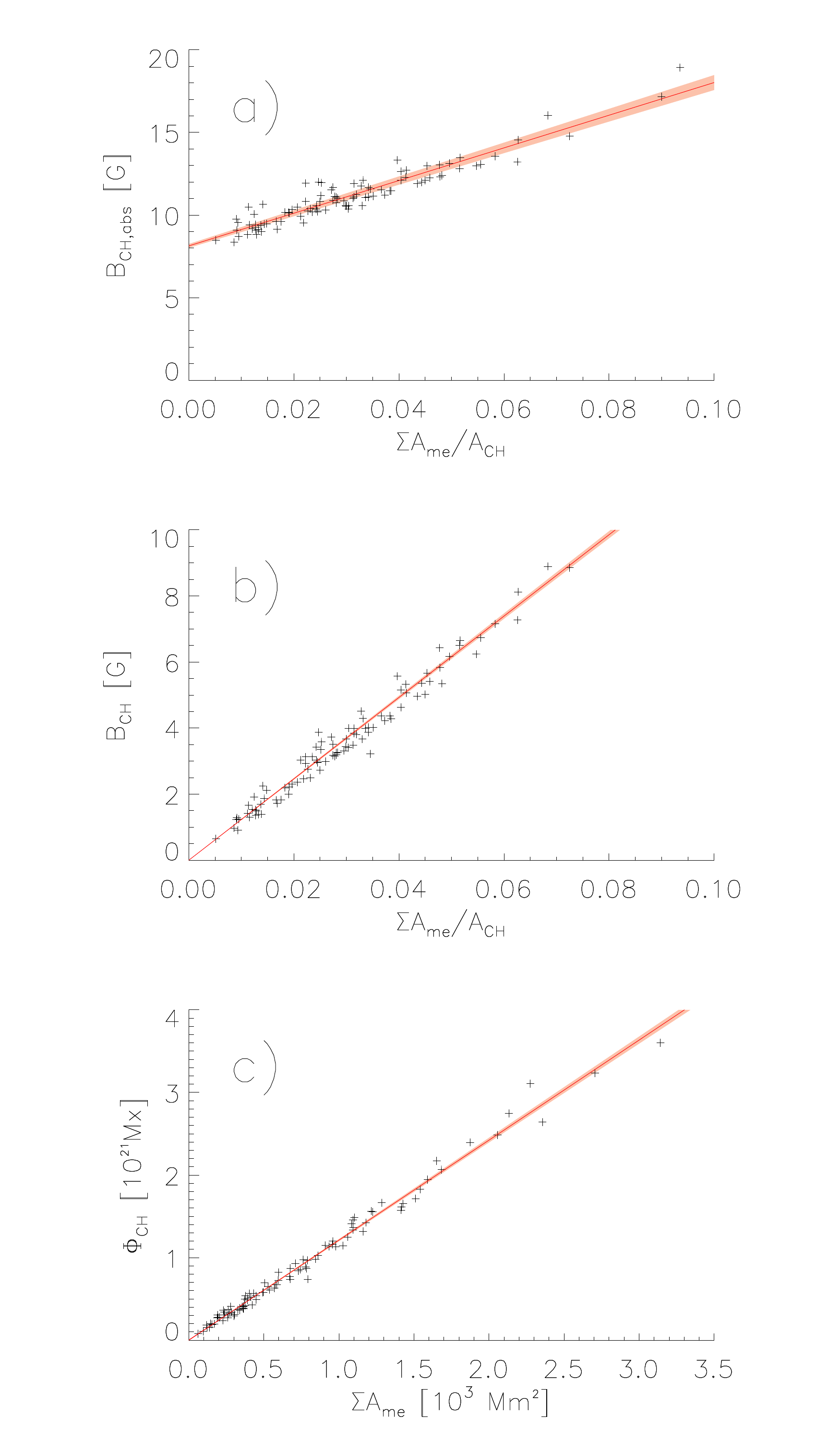}
\caption{Mean unsigned (a) and signed (b) magnetic field densities of the overall coronal holes versus their percentage coverage of long-lived magnetic elements. Unbalanced magnetic flux of the overall coronal holes versus their coverage of long-lived magnetic elements (c).  The shaded areas give the $1\sigma$ uncertainties of the corresponding fits.}
\label{consequences}
\end{figure}

Following this reasoning, the mean unsigned magnetic field strength of the overall coronal hole should be given by a constant offset representing the blurred noisy background magnetic field and the closed short- and medium-lived magnetic elements, and should further increase with increasing dominance of the long-lived magnetic elements. This is shown in Figure \ref{consequences}a. The offset at zero coverage with long-lived magnetic elements in the figure is \SI{8.1}{G}, which corresponds reasonably well to the unsigned magnetic flux density of \SI{7.1}{Mx/cm^2}, which stems from the quiet coronal hole regions due to magnetogram noise plus a contribution of \SI{0.7}{Mx/cm^2} from the short- and medium-lived magnetic elements (Table \ref{table_mc_stats}). Then, the mean unsigned magnetic field strength increases linearly with increasing percentage coverage of long-lived magnetic elements. The fit to the data is given by 
\begin{equation}
B_\text{CH,us} = \SI{8.1 \pm 0.1}{G} + \SI{99 \pm 3}{G} \cdot  \sum_\text{me} A_\text{me}/A_\text{CH},
\end{equation}
and the correlation coefficient is $0.95 \pm 0.01$.

The argumentation for the mean magnetic field strength is similar. The short- and medium-lived magnetic elements do not effectively contribute to the mean magnetic field strength; the main contribution arises from the small background magnetic field and the long-lived magnetic elements. However, since the strength of the background magnetic field is correlated with the average flux arising from the long-lived magnetic elements (Equ. \ref{cor_bg_long}), these two contributors can be merged. Then, the mean magnetic field strength simply linearly increases with increasing percentage coverage of the coronal hole with long-lived magnetic elements. This is shown in Figure \ref{consequences}b. The corresponding fit under the constraint that zero coverage of the magnetic elements results in no unbalanced magnetic flux for the overall coronal hole gives
\begin{equation}
B_\text{CH} = \SI{123 \pm 2}{G} \cdot  \sum_\text{me} A_\text{me}/A_\text{CH}. \label{consequences1}
\end{equation}
and follows perfectly the data. The correlation coefficient is $0.988 \pm 0.003$. 

Since the mean magnetic field strength is given by the percentage coverage of the coronal hole with long-lived magnetic elements, the total unbalanced magnetic flux of the coronal hole is given by the total area the long-lived magnetic elements cover. This is shown in Figure \ref{consequences}c. The  fit, again under the constraint that zero coverage results in no unbalanced magnetic flux, gives 
\begin{equation}
\Phi_\text{CH} = \SI{1.21 \pm 0.02 e18}{Mx} \cdot  \sum_\text{me} A_\text{me} /\si{Mm^2}. \label{eq_flux_area_con}
\end{equation}
The correlation coefficient $0.994 \pm 0.001$.
Again, we note that background magnetic field gives an essential contribution to the unbalanced magnetic flux of the coronal hole, which is already included in Equation \ref{eq_flux_area_con} because of its correlation with the long-lived magnetic elements.

\begin{figure}[tp]
\centering
\includegraphics[width = \linewidth]{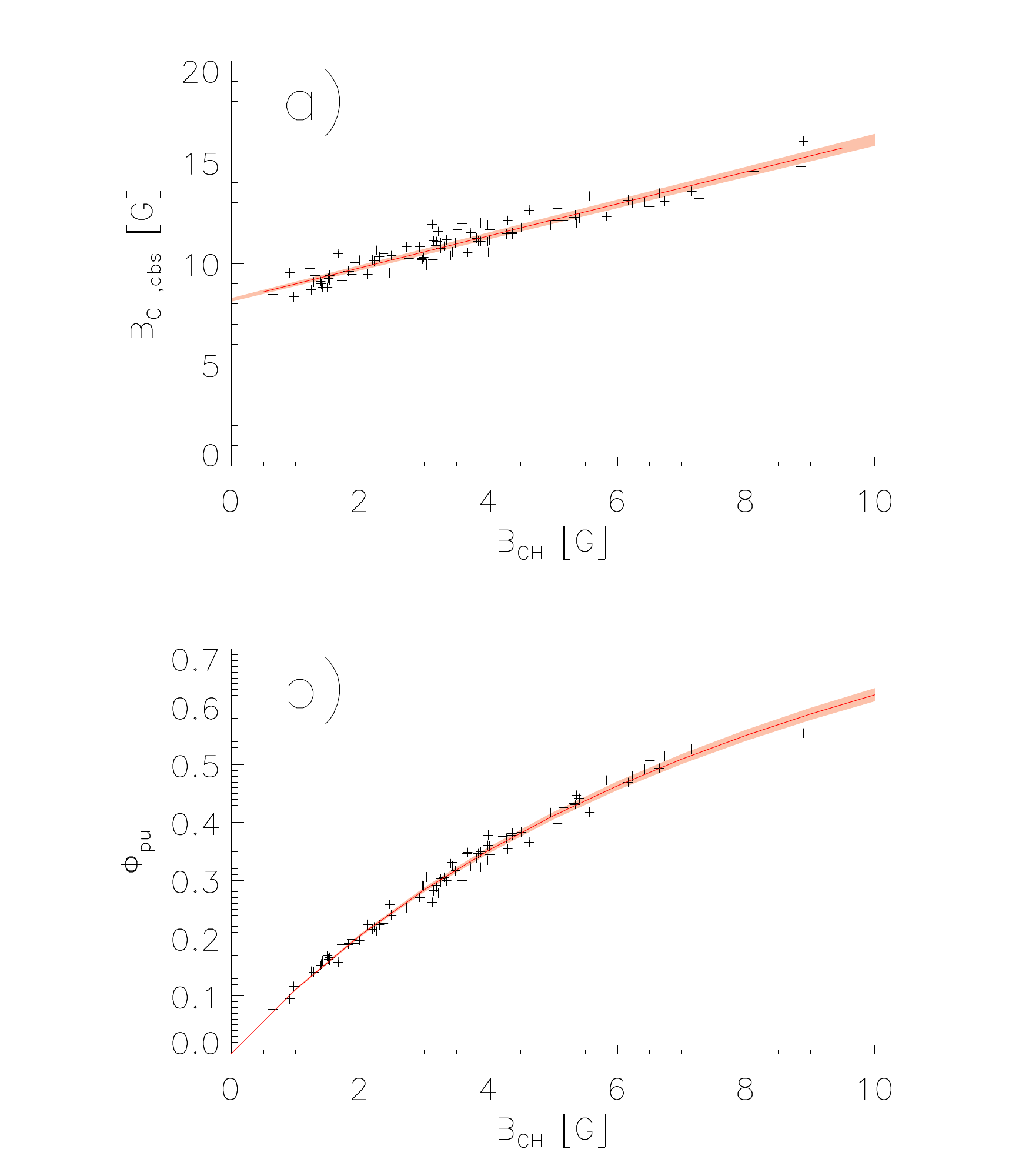}
\caption{Mean unsigned vs. mean signed magnetic field densities of the overall coronal holes, and the corresponding fit with its $1\sigma$ uncertainty (a). Percentage of the total unsigned magnetic flux of the coronal holes that does not close within the coronal hole boundaries vs. the mean magnetic field densities of the coronal holes. Equation \ref{eq_pu} with its $1\sigma$ uncertainty is overplotted in red (b).}
\label{consequences2}
\end{figure}

From the previous plots, two more important consequences arise. 
First, since both the mean unsigned magnetic field strength and the mean magnetic field strength are linearly increasing with increasing coverage of the coronal hole with long-lived magnetic elements, these two parameters are not independent. This is shown in Figure \ref{consequences2}a; the fit is 
\begin{equation}
B_\text{CH,us} = \SI{8.2 \pm 0.1}{G} + (0.79 \pm 0.02) \cdot B_\text{CH} \label{b_bus}
\end{equation}
at a correlation coefficient of $0.97 \pm 0.01$. This results from the fact that both parameters are a measure of the strength of the long-lived magnetic elements in the coronal hole; the only difference is that (1) the mean magnetic field strength contains a contribution from the background magnetic field, which is however correlated to the long-lived magnetic elements, and that (2) the mean unsigned magnetic field strength is further dependent on the local noise level in the magnetogram, which is however roughly constant within our dataset. 

Finally, the percentaged unbalanced magnetic flux of coronal holes, i.e., the percentage of the total unsigned magnetic flux of the coronal hole that does not close within the coronal hole boundary, is dependent on the mean magnetic field strength of the coronal hole. This follows from Equation \ref{b_bus} and the definition of the percentaged unbalanced magnetic flux, 
\begin{align}
\Phi_\text{CH,pu} &= \frac{\Phi_\text{CH}}{\Phi_\text{CH,us}} = \frac{B_\text{CH}}{B_\text{CH,us}}, \\
 &= \frac{B_\text{CH}}{\SI{8.2 \pm 0.1}{G} + (0.79 \pm 0.02) \cdot B_\text{CH}}. \label{eq_pu}
\end{align}
The relation together with Equation \ref{eq_pu} is shown in Figure \ref{consequences2}b; the Spearman's correlation coefficient of the data is $0.99 \pm 0.003$.
Basically, the percentaged unbalanced flux is meant to be a measure on how many of the magnetic funnels rooted in the magnetic elements close within the coronal hole boundaries and how many are open. However, since all the magnetic funnels having the long-lived magnetic elements as footpoints have the same polarity and are the main contributor to the unbalanced magnetic flux, all of these magnetic funnels can effectively be considered as open as long as the surrounding of the coronal hole is neglected. Further, the total unsigned magnetic field strength is dominated by magnetogram noise. Therefore, the percentaged unbalanced magnetic flux is only another measure on the number and strength of the open long-lived magnetic elements as compared to the noise-dominated background magnetic field distribution. It should not be interpreted as a measure on which percentage of the magnetic funnels belong to closed loop structures, respectively, open funnels. 

In summary, the number of open magnetic fibers given by the area the long-lived magnetic elements cover is the parameter that determines the mean magnetic field strength, the mean unsigned magnetic field strength, the unbalanced magnetic flux, and the percentaged unbalanced magnetic flux. The background magnetic field gives a small to minor contribution to the mean magnetic field strength that is correlated to the long-lived magnetic elements, and a major but constant contribution to the mean unsigned magnetic field strength dependent on the local noise level in the magnetogram. The unbalanced magnetic flux of the short- and medium-lived magnetic elements is negligible as compared to the long-lived magnetic elements and background magnetic field, and their contribution to the total unsigned flux is negligible as compared to the local noise level in the magnetogram.

\section{Discussion} \label{sec_discuss}
In this study, we investigated the structure of the photospheric magnetic field as observed with SDO/HMI LoS magnetograms and its consequences on the overall magnetic properties of coronal holes. Without the full-disk observations of AIA and HMI this study would not have been feasible. However, since HMI has a limited spatial resolution as compared to the actual state-of-the-art instruments such as Hinode/SP and ground-based observations, we dedicate this section to interpret our photospheric results in matters of high-resolution observations. Then, we extend in height and compare our photospheric results with coronal and solar wind observations.

\subsection{Photospheric magnetic structure}

Our main limitation is given by the spatial resolution of the HMI instrument of \SI{1.1}{arcsec}. When we observe photospheric structures that are much smaller than our resolution such as magnetic fibers, the limited spatial resolution leads to a strong underestimation of their actual magnetic flux densities and an overestimation of their size.

In our dataset, the quiet coronal hole regions consist mostly of pixels with magnetic flux densities of \SI{<<25}{G}. The real mean magnetic field densities of individual pixels in these regions is probably even less when we take the effect of magnetogram noise into account. These very low local magnetic field densities are partially an effect of the limited resolution of the HMI instrument. 
The quiet Sun is known to be covered by a high number of small-scale closed loops with footpoints in the inter-network of the granulational cells at typical magnetic field densities of \SIrange{100}{200}{G}, separated by voids of the magnetic field at the sizes of a few granules \citep{lites2008}. Since the orientation of these small-scale loops is mostly horizontal and their size is below the spatial resolution of the HMI instrument, their corresponding LoS signal is too small to be detected by HMI. 
Further, \citet{lites2008} also reported that small-scale vertical magnetic field regions appear all over the quiet Sun in the inter-network of the granulational cells. Their field strength is on the order of some \SI{100}{G}, i.e., at about the equipartition field strength. From the filling fractions of the corresponding pixels, which were estimated to be $\approx 0.2$ at Hinode's spatial resolution of \SI{\approx 0.3}{arcsec}, we can estimate their expected diameters to be on the order of \SI{0.13}{arcsec}. Therefore, these regions are too small to be identified by the HMI instrument, but they leave an imprint on the magnetic flux density distribution. Therefore, we might assume that the background magnetic flux density of \SIrange{0.2}{1.2}{G} in the quiet coronal hole regions we measured arises from a high number of these very small vertical magnetic fibers at the equipartition field strength. If these magnetic fibers further are the diffusive remnants of the long-lived magnetic elements, it would also explain the correlation between the flux of the background field and the flux of the long-lived magnetic elements. However, we also cannot fully exclude that the background magnetic field is an artefact owing to the point spread function (PSF) of the HMI instrument. The PSF could lead to a signal in the quiet coronal hole region that originally stems from the long-lived magnetic elements, and consequently to the observed correlation between the flux arising from the background magnetic field and arising from the long-lived magnetic elements.

The magnetic elements in our dataset are the source locations of assemblies of small-scale vertical magnetic fibers. When these magnetic fibers expand with height and when the directions of the magnetic fibers lies along the direction of the LoS, we can observe them as photospheric bright points. The magnetic field strength of these MBPs is reported to be in the range \SIrange{1.2}{1.5}{kG} \citep{utz2013} as determined by high-resolution observations. Therefore, these magnetic fibers have magnetic field strength far above the equipartion field strength of \SI{500}{G} and are thus a distinctly different phenomena than the unresolved magnetic fibers in the quiet coronal hole regions. The much lower magnetic flux densities of the magnetic elements in our dataset, on average \SI{\approx 50}{G}, as compared to the flux densities of the MBPs of \SIrange{1.2}{1.5}{kG}, is a result of the limited spatial resolution of HMI (\SI{1.1}{arcsec}) as compared to Hinode/SOT (\SI{0.2}{arcsec}): having only $1/5$ of the spatial resolution, the area per pixel increases and thus the mean magnetic flux density decreases by a factor of 25, leading to flux densities of \SI{50}{G} instead of \SI{1.25}{kG} per magnetic fiber.

The three classes of short- and medium-lived magnetic elements with lifetimes \SI{<40}{h} in our dataset are thought to be a result of convective motion or flows at granulational, meso-, and supergranulational scales. The class with the shortest half-life times of \SI{14}{min} match well the lifetimes of granules of \SIrange{7}{15}{min} \citep{spruit1990}, the second class with half-life times of \SI{2.1}{h} fits reasonably well the timescales of mesogranulation of \SI{\approx 3}{h} \citep{muller1992}, and the third class with a half-life of \SI{11.7}{h} corresponds well to the timescales of the supergranulation  of \SIrange{10}{36}{hours} \citep{rieutord2010}. The high magnetic field densities of their fibers of some \si{kG}, which are clearly above the equipartition magnetic field strength \SI{\approx 500}{G}, probably arise from convective collapse \citep{spruit1979, nagata2008, utz2014}.

The long-lived magnetic elements with lifetimes \SI{>40}{h}, on the other hand, look like the large magnetic elements of plage regions, i.e., remnants of the strong magnetic field of active regions. Almost all of these magnetic elements have the same dominant polarity of the coronal hole and they have extremely long lifetimes. Further, these magnetic elements are not at all rigid structures, but large clusters of individual magnetic fibers that can split and evolve independently. These long-lived magnetic elements are the main source of the unbalanced magnetic flux of low-latitude coronal holes. However, the old-standing question remains why they can be so long-lived without diffusing.

\subsection{Coronal hole structure}
Usually it is assumed that coronal holes appear dark owing to the outflowing solar wind, which reduces the temperature and density within the coronal hole. The plasma outflow itself was measured to be concentrated in magnetic funnels \citep{hassler1999, tu2005}, which are the vertical three-dimensional extensions of our photospheric magnetic elements. 
However, in Section \ref{sect_visual}, we have shown that the positions of the magnetic clusters and the darkest regions within the
coronal holes at coronal heights do not coincide, i.e., that the reduced extreme-ultraviolet (EUV) emission of coronal holes cannot simply be explained by the plasma outflow in purely vertical magnetic funnels.

This leaves us with two possibilities. Let us first presume that solar wind mainly arises from the magnetic funnels; this is the actual common believe. This would imply that the solar wind plasma is not flowing out of the darkest regions within coronal holes, and thus that dark regions and the solar wind plasma are not directly related.  Since the darkest coronal hole regions are located in the center of largely unipolar photospheric magnetic network cells, this would suggest that these regions are dark since there is almost no magnetic activity which heats them. 

The second possibility is that the solar wind does not only arise from the magnetic funnels, but also from the inner of the magnetic network cells. This would imply that the signal from the outflowing solar wind plasma in the magnetic network cells was simply too small to be measured by \citet{tu2005} with sufficient certainty. Such small signals have been reported by \citet{aiouaz2005}. In this case, an essential part of the high-speed solar wind plasma could also stem from the inner of the magnetic network cells owing to its much larger area as compared to the total area of the magnetic funnels.

In each of the cases, it seems as there is an important piece missing that relates the low EUV intensity of coronal holes to its photospheric magnetic field configuration. Just assuming that coronal holes appear dark because of the outflowing solar wind concentrated in vertical magnetic funnels is probably an insufficient interpretation.

\subsection{High-speed solar wind streams}

\citet{wang2010} and \citet{wang2013} derived that the mass flux of high-speed solar wind streams close to the Sun depends on the mean magnetic field strength of their source regions by performing a back-calculation on solar wind in situ data measured close to the Earth and applying conservation of mass flux. In particular, they derived 
\begin{equation}
n_0 v_0 \approx \SI{8e12}{cm^{-2} s^{-1}} \cdot B_\text{CH} /\si{G}
\end{equation}
for the time range 1998 to 2012. Using Equation \ref{consequences1}, we can easily explain this dependence. A higher mean magnetic field strength is a result of a higher coverage of the coronal hole with magnetic elements of all magnitudes. These, on the other hand, are thought to be the small-scale source regions of the solar wind. Thus, a higher mean magnetic field strength of the coronal hole leads to a higher coverage with small-scale solar wind sources within the coronal hole, and thus to a higher average solar wind mass flux at large scales.

\section{Conclusions} \label{sec_conclude}

In this study, we have statistically analyzed the magnetic field below coronal holes and related it to the overall characteristics of coronal holes.
We found that 
\begin{enumerate}
\item The long-lived magnetic elements with lifetimes \SI{>40}{hours} are the main source of the unbalanced magnetic flux of coronal holes. They have mostly the dominant polarity of the coronal hole and account for on average \SI{69 \pm 8}{\percent} of the unbalanced magnetic flux of coronal holes. Further, the area they cover completely determines the mean magnetic flux strength and unbalanced magnetic flux of the coronal hole ($cc= 0.988 \pm 0.003$), with
\begin{align*}
B_\text{CH} &= \SI{123 \pm 2}{G} \cdot  \sum_\text{me} A_\text{me}/A_\text{CH},\\
\Phi_\text{CH} &= \SI{1.21 \pm 0.02e18}{Mx} \cdot  \sum_\text{me} A_\text{me} / \si{Mm^2}.
\end{align*}

\item The shorter-lived magnetic elements with lifetimes \SI{<40}{hours} can be related to granulation, meso-granulation, and super-granulation as based on their lifetimes. They appear with both polarities at about similar rates; their total contribution to the unbalanced magnetic flux is only \SI{\approx 5.0 \pm 0.1}{\percent}. They are likely mostly the footpoints of small-scale closed loops.

\item The photosphere below coronal holes outside of the magnetic elements, i.e., the quiet coronal hole regions, contribute \SI{\approx 22 \pm 4}{\percent} to the unbalanced magnetic flux of coronal holes. The magnetic flux can be well modeled as a background magnetic field strongly blurred by noise with a second, smaller contribution from probably unresolved magnetic elements. The total unbalanced magnetic flux from this region is positive correlated to the magnetic flux arising from the long-lived magnetic elements ($cc=0.88 \pm 0.02$).

\item The location of the photospheric magnetic elements do not correspond to the darkest coronal regions within the coronal holes. Therefore, the darkness of coronal holes cannot be easily explained by plasma outflow of vertical magnetic funnels that are rooted in the magnetic elements.

\item All magnetic elements inside coronal holes follow a distinct area-flux relationship ($cc = 0.984 \pm 0.001$), given by a power law (Eq. \ref{eq_flux_area_mc})
\begin{equation*}
\Phi_\text{me} = \SI{0.406 \pm 0.002e18}{Mx} \cdot \left( A_\text{me} / \si{Mm^2} \right)^{(1.261 \pm 0.004)},
\end{equation*}
independent of their lifetime.
\item The magnetic elements show a substructure, magnetic fibers, as outlined by the individual photospheric MBPs observed in the G band. Therefore, the magnetic elements are not rigid objects; they can split and evolve separately. The largest magnetic element in our dataset contained 51 MBPs.

\end{enumerate}

We conclude that main building block of the strong magnetic field of coronal holes are small-scale magnetic fibers that are visible as photospheric MBPs in G-band filtergrams. They appear mostly clustered at the edges of the supergranular network, and their enclosing contours are observable as magnetic elements in photospheric magnetograms. Because of that, the total area the magnetic elements cover is an estimate of the number of magnetic fibers within the coronal hole and, presuming that the magnetic flux of the magnetic fibers is statistically distributed, a measure of their total magnetic flux. Since the short-lived magnetic elements are well balanced in their opposite fluxes and only the long-lived magnetic elements have a common polarity, only these contribute significantly to the overall unbalanced magnetic flux. Following from this, a higher coverage with long-lived magnetic elements of the coronal hole is synonymous with having a larger number of strong magnetic fibers of the dominant polarity, which results in a larger amount of unbalanced magnetic flux of the overall coronal hole.

The second, smaller contributor to the unbalanced magnetic flux is the quiet coronal hole region, i.e., the region outside of the magnetic elements within the coronal hole. We found that this magnetic flux likely arises from a background magnetic field in the range \SIrange{0.2}{1.2}{G,} which is strongly blurred by noise, and not from unresolved magnetic elements. Since its flux is well correlated to the magnetic flux from the long-lived magnetic elements, there has to be a mechanism that couples these two. 

Following the results that the unbalanced magnetic flux arises from the long-lived magnetic elements and a background magnetic field, this also means that the magnetic field of a coronal hole can be modeled as a weak background magnetic field in which a small number of strong magnetic elements are set into the supergranular network lanes. When the artificial magnetic elements follow the distinct area-flux relationship as given by Equation \ref{eq_flux_area_mc} and are randomly drawn from it, their total area gives the total unbalanced magnetic flux of the coronal hole. Then, according to Section \ref{sec_overall}, the overall magnetic properties and dependences of the coronal hole between the mean magnetic field strength, mean unsigned magnetic field strength, unbalanced magnetic flux, and percentaged unbalanced magnetic flux we observe will be automatically correctly fulfilled.

Finally, the question arises whether the photospheric magnetic field below coronal holes is unique. A perfect answer to this question is probably not possible. We know that the unbalanced magnetic flux of coronal holes mainly arises from the long-lived magnetic elements, and that these magnetic elements look like active region plage. Since plage regions exist not only below coronal holes but all over the Sun, we believe that the photospheric magnetic field below coronal hole is not unique. If their photospheric magnetic field is not unique, this means that coronal holes probably do not have their origin in the solar interior since we should see a footprint of it in the photosphere. Then the reason why coronal holes exist has to be either the global magnetic field configuration leading to an open magnetic field configuration, or the plasma outflow in the magnetic funnels, which opens up and stabilizes the local open magnetic field configuration.

\begin{acknowledgements}
We thank the referee for a careful reading of the manuscript and his constructive and detailed comments, which clearly helped to improve this comprehensive study.
The SDO/AIA images and SDO/HMI images are available by courtesy of NASA/SDO and the AIA, EVE, and HMI science teams.
Hinode is a Japanese mission developed and launched by ISAS/JAXA, with NAOJ as domestic partner and NASA and STFC (UK) as international partners. It is operated by these agencies in cooperation with ESA and NSC (Norway). The Hinode/SOT images are available by the courtesy of the Hinode science team. 
S. Hofmeister thanks the JungforscherInnenfond der Steierm\"arkischen Sparkassen for their support.
D. Utz and A.M. Veronig acknowledge the support by the Austrian Science Fund (FWF) P27800 and P24092-N16.
M. Temmer, S.G. Heinemann, and A.M. Veronig acknowledge the support by the FFG/ASAP Program under grant No. 859729 (SWAMI).
\end{acknowledgements}

\end{document}